\documentclass[10pt,journal,compsoc]{IEEEtran}
\usepackage{graphicx}
\usepackage{array}
\usepackage{url}
\usepackage{fancybox}
\usepackage{multirow}
\usepackage{amssymb}
\usepackage{bigstrut}
\usepackage{color}

\usepackage{colortbl}
\usepackage{balance}
\usepackage{fancyhdr}
\usepackage{cite}
\usepackage{subfig}
\usepackage{diagbox}
\usepackage{bbding}
\usepackage{placeins}

\usepackage{booktabs}
\usepackage{enumitem}
\usepackage{xcolor,lipsum}
\usepackage{listings}
\usepackage{xcolor}
\usepackage{times}
\usepackage{balance}

\begin{document}
	\title{Defining Smart Contract Defects on Ethereum}
	
	\author{Jiachi Chen, Xin Xia, David Lo, John Grundy, Xiapu Luo and Ting Chen
	\IEEEcompsocitemizethanks{\IEEEcompsocthanksitem Jiachi Chen, Xin Xia and John Grundy are with the Faculty of Information Technology, Monash University, Melbourne, Australia. \protect\\
		E-mail: \{Jiachi.Chen, Xin.Xia, John.Grundy\}@monash.edu
		\IEEEcompsocthanksitem David Lo is with the School of Information Systems, Singapore Management University, Singapore.\protect\\
		E-mail: davidlo@smu.edu.sg
		\IEEEcompsocthanksitem Xiapu Luo is with the Department of Computing, The Hong Kong Polytechnic University, Hong Kong.\protect\\
		E-mail: csxluo@comp.polyu.edu.hk
		\IEEEcompsocthanksitem Ting Chen is with the School of Computer Science and Engineering, University of Electronic Science and Technology of China, China.\protect\\
		E-mail: brokendragon@uestc.edu.cn
		\IEEEcompsocthanksitem Xin Xia is the corresponding author.}
	\thanks{Manuscript received     ; revised   }}

\markboth{IEEE Transactions on Software Engineering, ~Vol.~  , No.~  , 2020 }%
{Shell \MakeLowercase{\textit{et al.}}: Bare Demo of IEEEtran.cls for Computer Society Journals}

	\IEEEtitleabstractindextext{%
	\begin{abstract}
		\emph{Smart contracts} are programs running on a blockchain. They are immutable to change, and hence can not be patched for bugs once deployed. Thus it is critical to ensure they are bug-free and well-designed before deployment. A \emph{Contract defect} is an error, flaw or fault in a smart contract that causes it to produce an incorrect or unexpected result, or to behave in unintended ways. The detection of contract defects is a method to avoid potential bugs and improve the design of existing code. Since smart contracts contain numerous distinctive features, such as the \textit{gas system. decentralized}, it is important to find smart contract specified defects. To fill this gap, we collected smart-contract-related posts from Ethereum StackExchange, as well as real-world smart contracts. We manually analyzed these posts and contracts; using them to define 20 kinds of \emph{contract defects}. We categorized them into indicating potential security, availability, performance, maintainability and reusability problems. To validate if practitioners consider these contract  as harmful, we created an online survey and received 138 responses from 32 different countries.  Feedback showed these contract defects are harmful and removing them would improve the quality and robustness of smart contracts. We manually identified our defined contract defects in 587 real world smart contract and publicly released our dataset. Finally, we summarized 5 impacts caused by contract defects. These help developers better understand the symptoms of the defects and removal priority.
		
	\end{abstract}
	
	\begin{IEEEkeywords}
		Empirical Study, Smart Contracts, Ethereum, Contract Defect
\end{IEEEkeywords}}
	\maketitle
\IEEEdisplaynontitleabstractindextext

	%
	%
	%
	%
	%
	%

	\section{Introduction}
\label{Introduction}
The considerable success of decentralized cryptocurrencies has attracted great attention from both industry and academia. Bitcoin~\cite{bitcoin} and Ethereum~\cite{whitepaper,yellowpaper} are the two most popular cryptocurrencies whose global market cap reached \$162 billion by April 2018~\cite{marketcap}. A \emph{Blockchain} is the underlying technology of cryptocurrencies, which runs a consensus protocol to maintain a shared ledger to secure the data on the blockchain. Both Bitcoin and Ethereum allow users to encode rules or scripts for processing transactions. However, scripts on Bitcoin are not Turing-complete, which restrict the scenarios of its usage. Unlike Bitcoin, Ethereum provides a more advanced technology named \emph{Smart Contracts}.

Smart contracts are Turing-complete programs that run on the blockchain, in which consensus protocol ensures their correct execution~\cite{whitepaper}. With the assistance of smart contracts, developers can apply blockchain techniques to different fields like gaming and finance. When developers deploy smart contracts to Ethereum, the source code of contracts will be compiled into \textit{bytecode} and reside on the blockchain. Once a smart contract is created, it is identified by a 160-bit hexadecimal address, and anyone can invoke this smart contract by sending transactions to the corresponding contract address. Ethereum uses \textit{Ethereum Virtual Machine (EVM) }to execute smart contracts and transaction are stored on its blockchain.

A blockchain ensures that all data on it is immutable, i.e., cannot be modified, which means that smart contracts cannot be patched when bugs are detected or feature additions are desired. The only way to remove a smart contract from blockchain is by adding a \textit{selfdestruct}~\cite{Solidity} function in their code. 
Even worse, smart contracts on Ethereum operate on a permission-less network. Arbitrary developers, including attackers, can call the methods to execute the contracts. For example, the famous\textit{\textbf{ DAO attack}}~\cite{DAOAttack} made the DAO (Decentralized Autonomous Organization) lose 3.6 million Ethers (\$150/Ether on Feb 2019), which then caused a controversial hard fork~\cite{hardfork, Etc} of Ethereum. 

It is thus critical to ensure that smart contracts are bug-free and well-designed before deploying them to the blockchain. In software engineering, a software defect is an error, flaw or fault in a computer program or system that causes it to produce an incorrect or unexpected result, or to behave in unintended ways~\cite{defect_wiki, chillarege1996orthogonal}. Contract defects are related to not only security issues but also design flaws which might slow down development or increase the risk of bugs or failures in the future. Detecting and removing contract defects helps increase software robustness and enhance development efficiency~\cite{van2002java,khomh2009exploratory}. Since the revolutionary changes of smart contracts compared to traditional softwares, e.g., the gas system, decentralized features, smart contracts contain many specific defects. 



In this paper, we conduct an empirical study on defining smart contracts defects on Ethereum platform, the most popular decentralized platform that runs smart contracts. Please note that some previous works~\cite{oyente, Maian, Zeus} focus on improving the quality of smart contracts from the security aspect. However, this is the first paper that aims to provide a systematic study of contract defects from five aspects: security, availability, performance, maintainability and reusability. These previous works were not comprehensive and did not validate whether practitioners consider these contract defects as harmful. To address these limitations, we conducted our results from 17,128 \textit{Ethereum.StackExchange}\footnote{https://ethereum.stackexchange.com/} posts and validated it by an online survey. To help developers better understand the symptoms and distribution of smart contract defects, we manually labeled a dataset and released it publicly to help further study. In this paper, we address the following key research questions:

\noindent\textbf{\textit{RQ1: What are the smart contract defects in Ethereum?}}

We identified and defined 20 smart contract defects from \textit{StackExchange} posts and real-world smart contracts. These 20 contract defects are considered from security, availability, performance, maintainability and reusability aspects. By removing the defined defects from the contracts, it is likely to improve the quality and robustness of the programs.

\noindent\textbf{\textit{RQ2: How do practitioners perceive the contract defects we identify?}}

To validate the acceptance of our newly defined smart contract defects, we conducted an online survey and received 138 responses and 84 comments from developers in 32 countries. The options in the survey are from 'Very important' to 'Very unimportant' and we give each option a score from 5 to 1, respectively. The average score of each contract defect is 4.22. The feedbacks and comments show that developers believe removing the defined contract defects can improve the quality and robustness of smart contracts.

\noindent\textbf{\textit{RQ3: What are the distributions and impacts of the defects in real-world smart contracts?}}

We manually labeled 587 smart contracts and found that more than 99\% of smart contracts contain at least one of our defined defects. We then summarized 5 impacts that can help researchers and developers better understand the symptoms of these contract defects.

The main contributions of this paper are:

\begin{itemize}
	\item We define 20 contract defects for smart contracts considering five aspects: \emph{security, availability, performance, maintainability} and \emph{reusability}.  We list symptoms and give a code example of each contract defects, which can help developers better understand the defined contract defects. To help further researches, we also give possible solution and possible tools for the contract defects. 
	
	\item We manually identify whether the defined 20 defects exist in real-life smart contracts. Our dataset\footnote{The dataset can be found at https://github.com/Jiachi-Chen/TSE-ContractDefects} contains a collection of 587 smart contracts, which can assist future studies on smart contract analysis and testing. Also, we analyze the impacts of the defined contract defects and summarize 5 common impacts. These impacts can help developers decide the priority of defects removal.
	
	
	\item Our work is the first empirical study on contract defects for smart contracts. We aim to identify their importance, and gather inputs from practitioners. This work is a requirement engineering step for a practical contract defects detection tool, which is an important first step that can lead to the development of practical and impactful tools to practitioners.
	
\end{itemize}

The remainder of this paper is organized as follows. In Section 2, we provide background knowledge of smart contracts. In Sections 3-5, we present the answers to the three research questions, respectively. We discuss the implications,  and challenge in automatic contract defects detection in Section 6.  In Section 7, we introduce threats to validity. Finally, we elaborate the related work in Section 8, and conclude the whole study and mention future work in Section 8.

\section{Background}
\label{background}
In this section, we briefly introduce background knowledge about smart contracts as well as the Solidity programming language for smart contract definition. 
\subsection{Smart Contracts - A Decentralized Program}

A smart contract is \emph{``a computerized transaction protocol that executes the terms of contract"}  ~\cite{tapscott2016blockchain}. Their bytecode and transactions are all stored on the blockchain and visible to all users. Since Ethereum is an add-only distributed ledger, once smart contracts are deployed to a blockchain, they are immutable to be modified even when bugs are detected. Once a smart contract is created, it is identified by a unique 160-bit hexadecimal string referred to as its contact address. The Ethereum Virtual Machine (EVM) is used to run smart contracts. The executions of smart contracts depend on their code. For example, if a contract does not contain functions that can transfer Ethers, even the creator can not withdraw the Ethers. Once smart contracts are deployed, they will exist as long as the whole network exists unless they execute \emph{selfdestruct} function~\cite{Solidity}. \emph{selfdestruct} is a function that if it is executed, the contract will disappear and its balance will transfer to a specific address. In this paper, we describe smart contracts developed using \emph{Solidity}~\cite{Solidity}, the most popular smart contract programming language in Ethereum.

\subsection{Features of Smart Contracts}
\noindent {\bf The Gas System.}
In Ethereum, miners run smart contracts on their machines. As compensation for miners who contribute their computing resources, the creators and users of smart contracts will pay a certain amount of Ethers to the miners. The Ethers that are paid to miners are computed by: \textit{gas cost} * \textit{gas price}. Gas cost depends on the computational resource the transaction will take and gas price is offered by the transaction creators. The minimum unit of gas price is \emph{Wei} (1 Ether = $10^{18}$ Wei). The miners have the right to choose which transaction can be executed and broadcasted to the other nodes on the blockchain ~\cite{yellowpaper}. Therefore, if the gas price is too low, the transactions may not be executed. To limit the gas cost, when a user sends a transaction to invoke a contract, there will be a limit (\emph{Gas Limit}) that determines the maximum gas cost. If the gas cost exceeds the \emph{Gas Limit}, the execution is terminated with an exception often referred to as  \emph{out-of-gas error}.

\noindent {\bf Data location.}
In smart contracts, data can be stored in  \emph{storage}, \emph{memory} or \emph{calldata}~\cite{Solidity}. \emph{storage} is a persistent memory area to store data. For each \emph{storage variable}, EVM will assign a storage slot ID to identify it. Writing and reading \emph{storage variable} is the most expensive operation as compared with reading from the other two locations. The second memory area is named \emph{memory}. The data of the \emph{memory variables} will be released after their life cycle finished. Writing and reading to \emph{memory} is cheaper than \emph{storage}. \textit{Calldata} is only valid for parameters of external contract functions. Reading data from the \emph{Calldata} is much cheaper than \emph{memory} or \emph{storage}.

\subsection{Solidity}
Solidity is the most popular programming language that is used to program smart contracts on the Ethereum platform. In this subsection, we give a basic overview of Solidity programming as well as a Solidity example.

\noindent {\bf Fallback Function.}
The fallback function~\cite{Solidity} is the only unnamed function in Solidity programming. This function does not have arguments or return values. It is only executed when an error function call happens. For example, a user calls function ``$\delta$" but the callee contract does not contain this function. The fallback function will be executed to handle the error. Also, if a fallback function is marked by \emph{payable}\footnote{If a function wants to receive Ethers, it has to add \emph{payable}}, e.g., line 13 in listing 1, it will be executed automatically when the contract receives Ethers.

\begin{lstlisting}[caption={A ``Gamble" smart contract. However, this contract contains several contract defects.}]
pragma solidity ^0.4.25;
contract Gamble{
  address owner;
  address[] members;
  address[] participators;
  uint participatorID = 0;
  modifier onlyOwner{/*Transaction State Dependency*/
	require(tx.origin==owner); 
	 _; }
  function constructor(){ //constructor_function
	owner = //this is the address of tx.origin
		0xdCad...d1D3AD; /*Hard Code Address*/}
  function() payable{ //Executed when receiving Ethers
	ReceiveEth();}
  function ReceiveEth() payable{
	if(msg.value!=1 ether){
	  revert();}//msg.value is the number of received ETHs
	members.push(msg.sender);
	participators[participatorID] = msg.sender;
	participatorID++;
	if(this.balance==10 ether){/*Strict Balance Equality*/
  	  getWinner();}}
  function getWinner(){ //choose a member to be the winner
	/*Block Info Dependency*/
	uint winnerID = uint(block.blockhash(block.number)) % participants.length;
	participants[winnerID].send(8 ether);
	participatorID = 0;}
  function giveBonus() returns(bool){ //send 0.1 ETH to all members as bonus
	/*Unmatched Type Assignment, Nested Call*/
	for(var i = 0;i < members.length; i++){
	  if(this.balance > 0.1 ether)
		/*DoS Under External Influence*/
		members[i].transfer(0.1 ether); }
		/*Missing Return Statement*/ }
  function suicide(address addr) onlyOwner{ //Remove the contract from blockchain
	selfdestruct(addr);}
  function withDraw(uint amount) onlyOwner{ //withdraw certain Ethers to owner account
	address receiver = 0x05f4...d27;
	receiver.call.value(amount);}}
\end{lstlisting}

\noindent {\bf Ether Transfer and Receive.}
Solidity provides three APIs to transfer Ethers between accounts, i.e., \textit{address.transfer(amount), address.send(amount)}, and \textit{address.call.value(amount)()}. \textit{transfer} and \textit{send} will limit the gas of fallback function in callee contracts to 2300 gas~\cite{Solidity}. This gas is not enough to write to storage, call functions, or send Ethers. Therefore, \emph{transfer} and \emph{send} functions can only be used to send Ethers to \textit{External Owned Accounts} (EOA). \footnote{There are two types of accounts on Ethereum: externally owned accounts which controlled by private keys, and contract accounts which controlled by their contract code.} \textit{call} will not limit the gas of fallback function. Therefore,  \textit{call} can be used to send Ethers to either contract or EOA. The difference between \textit{transfer} and \textit{send} is that \textit{transfer} will throw an exception and terminate the transaction if the Ether fails to send, while  \textit{send} will return a boolean value instead of throwing an exception.

\noindent {\bf Version Controller.}
Ethereum supports multiple versions of Solidity. When deploying a smart contract to the Ethereum, developers need to choose a specific Solidity compiler version to compile the contract. Solidity is a young and evolving programming language. There are more than 20 versions released up to 2019. Different versions might have several significant language changes. If developers do not choose the correct version of Solidity, the smart contract compilation might fail. To make code reuse easier, a contract can be annotated with \textit{version pragma} that indicates the version that supported. The version pragma is used as: ``\emph{pragma solidity $\hat{ }$version}" or ``\emph{pragma solidity version}". For example,  ``\emph{pragma solidity $\hat{ }$0.4.1}" means that this contract supports compile version 0.4.1 and above (except for v0.5.0) while ``\emph{pragma solidity 0.4.1}" means that the contract only supports compile version 0.4.1. 

\noindent {\bf Permission Check.}
Smart contracts on Ethereum run in a permission-less network; everyone can call methods to execute the contracts. Developers usually add permission checks for permission-sensitive functions. For example, the contract will record the owner’s address in its constructor function as the constructor function can only be executed once when deploying the contract to the blockchain. In each transaction, the contract compares whether the caller’s address is the same as the owner’s address. Solidity provides \textit{msg} related APIs to receive caller information. For example, contracts can get the caller address from \textit{msg.sender}. Besides, Solidity also provides function modifiers to add prerequisite checks to a function call. A function with a function modifier can be executed if it passes the check of the modifier. 

\noindent {\bf Solidity Example.}
Listing 1 is a simple example of a smart contract which is developed in Solidity.  The contract is a gambling contract, each gambler sends 1 Ether to this contract. When the contract receives 10 Ethers, it will choose one gambler as the winner and sends 8 Ethers to him. 

The first line indicates the contract supports compiler version 0.4.25 to 0.5.0 (not included). The \textit{modifier} on line 7 is used to restrict the behavior of functions. For example,  \textit{onlyOwner} requires the \textit{tx.origin} equals to the \textit{owner}, and tx.origin is used to get the original address that kicked off the transaction, otherwise, the transaction will be roll back. If a function contains modifiers the function will first execute the modifiers. Line 10 is the constructor function of the contract. This function can only be executed once when deploying the contract to Ethereum. In the constructor function, the contract assigns a hard-coded address to the \textit{owner} variable to restore the owner address. Fallback function (L13) is a specific feature in smart contract as we introduced in Section 2.2. When receiving Ethers, \textit{ReceiveEth} will be activated and the contract uses \textit{msg.value} to check the amount of Ethers they received (L16). If the amount that they received not equal to 1 Ether, the transaction will be reverted. Otherwise, the contract records the address of those who send the Ethers (L18). When the balance equals to 10 Ethers, the contract will execute the \textit{getWinner} function and choose one gambler as the winner (L21-22). The function uses \textit{block.hash} and \textit{block.number} to generate a random number. This two block-related APIs is used to obtain block related information. After getting the winner, the contract uses \textit{address.send()} to send Ethers to the winer  (L26). \textit{address.send()} is one method to send Ethers. This method will return a boolean value to inform the caller whether the money is successfully sent but do not throw an exception. \textit{address.transfer()} can also be used to send Ethers, but this function will throw an exception when errors happen. Note that, these two functions have \textit{gas limitation} of 2300 if the recipient is a contract account (See Section 2.2).  \textit{address.call.value()} in Line 39 can be used to send Ethers to a smart contract, similar to \textit{address.send()}. This method also returns a boolean value to inform the caller whether the money is successfully sent but does not throw an exception.

\subsection{ERC-20 Token}
In recent years, thousands of cryptocurrencies have been created. However, most of them are implemented by smart contracts that run on the Ethereum (also called tokens) rather than having their own blockchain system. Ethereum provides several token standards to standardize tokens’ behaviors. In this case, different tokens can interact accurately and be reused by other applications (e.g., wallets and exchange markets). The ERC-20 standard~\cite{erc20} is the most popular token standard used on Ethereum. It defines 9 standard interfaces (3 are optional) and 2 standard events. To design ERC-20 compliant tokens, developers must strictly follow this standard. For example, the standard method \textit{transfer} is declared as ``function transfer (address \_to, uint256 \_value) public returns (bool success)”, which is used to transfer a number of tokens to address
\textit{\_to}. The function should fire the \textsc{TRANSFER} event to inform whether the tokens are transferred successfully. The function also should \textit{throw} an exception if the message caller’s account balance does not have enough tokens to spend.


\section{RQ1: Contract defects in Smart Contracts}
\label{definition}
\subsection{Motivation}
Smart contracts cannot be patched after deploying them to the blockchain. Detecting and removing contract defects is a good way to ensure contacts' robustness. Since the revolutionary changes of smart contracts compared to traditional softwares, e.g., the gas system, decentralized features, smart contracts might contain many specific defects compared to traditional programs, e.g., Android Apps. To fill this gap, we try to define a set of new smart contract defects from \textit{StackExchange} posts in this section. We give definitions, examples and possible solutions of our defined contract defects specialized for Ethereum smart contracts.

\subsection{Approach}

\noindent \textit{\textbf{3.2.1. StackExchange Posts}}: To define defects for smart contracts, we need to collect issues that developers encountered. Programmers often collaborate and share experience over Q\&A site like \emph{Ethereum StackExchange}~\cite{StackExchange}, the most popular and widely-used question and answer site for users of Ethereum. By analyzing posts on \emph{Ethereum StackExchange}, we can identify and define a set of contract defects on Ethereum. In this paper, we crawled 17,128 \textit{StackExchange} posts and analyzed them further. 

\noindent 	\textit{\textbf{3.2.2. Key Words Filtering}}: It is time-consuming to find important information from thousands of Q\&A posts. Therefore, we utilized keywords to filter important information from StackExchange posts. To ensure the completeness of our keywords list, two authors of this paper read the solidity documents~\cite{Solidity} carefully and recorded the keywords they think are important. After that, they merged the keywords list and used these keywords to filter StackExchange posts. When reading the posts, we added new keywords to enrich our list and filter new posts. We finally used 66 keywords to filter 4,141 posts.

\noindent \textit{\textbf{3.2.3. Manually Filtering}}: In this paper, we aim to find Solidity-related smart contract defects. However, the filtered 4,141 posts which contain the keywords might not related to Solidity-related contract defects. Many posts are about the web3~\cite{web3}, development environment (Remix~\cite{remix}, Truffle~\cite{truffle}), wallet or functionality. We need to remove them from the dataset and only retain posts that are related to contract defects. For example, the title of a post is ``Transfer ERC20 token from one account to another using web3”. Although the post contains key words ``ERC20”, the posts are related to web3, not Solidity related contract defects. Therefore, we emit it from our dataset. Two authors of this paper, who both have rich experience in smart contract development, manually analysed all of the posts and finally found that a total of 393 posts are related to Solidity-related smart contract defects. The detailed analysis results of these 4,141 posts can be found at: https://github.com/Jiachi-Chen/TSE-ContractDefects

\begin{table}
	\scriptsize
	\caption{Classification scheme.}
	\label{tab:classification}
	\centering
	\begin{tabular}{p{55pt} | p{150pt}}
		\hline
		Category & Description\\
		\hline
		Gas Limitation & Bugs caused by gas limitation. \\
		\hline
		Permission Check & Bugs caused by permission check failure.\\
		\hline
		Inappropriate Logic & There are inappropriate logics inside a contract, which can be utilized by attackers.\\
		\hline
		Ethereum Features & Ethereum has many new features, e.g., Solidity, Gas System. Developers do not familiar with the differences which might lead to mistakes. \\
		\hline
		Version Gaps & Errors due to the update of Ethereum or Solidity. \\
		\hline
		Inappropriate Standard & Ethereum provides several standards, but many contracts do not follow them.  \\
		\hline
	\end{tabular}	
\end{table}

\noindent \textit{\textbf{3.2.4. Open Card Sorting}}: We followed the card sorting~\cite{cardsort} approach to analyze and categorize the filtered contract defects-related posts. We created one card for each post. The card contains the information of defect title, description, and comments. The same two authors worked together to determine the labels of each post. The detailed steps are:
\begin{figure}
	\begin{center}
		\includegraphics[width=0.49\textwidth]{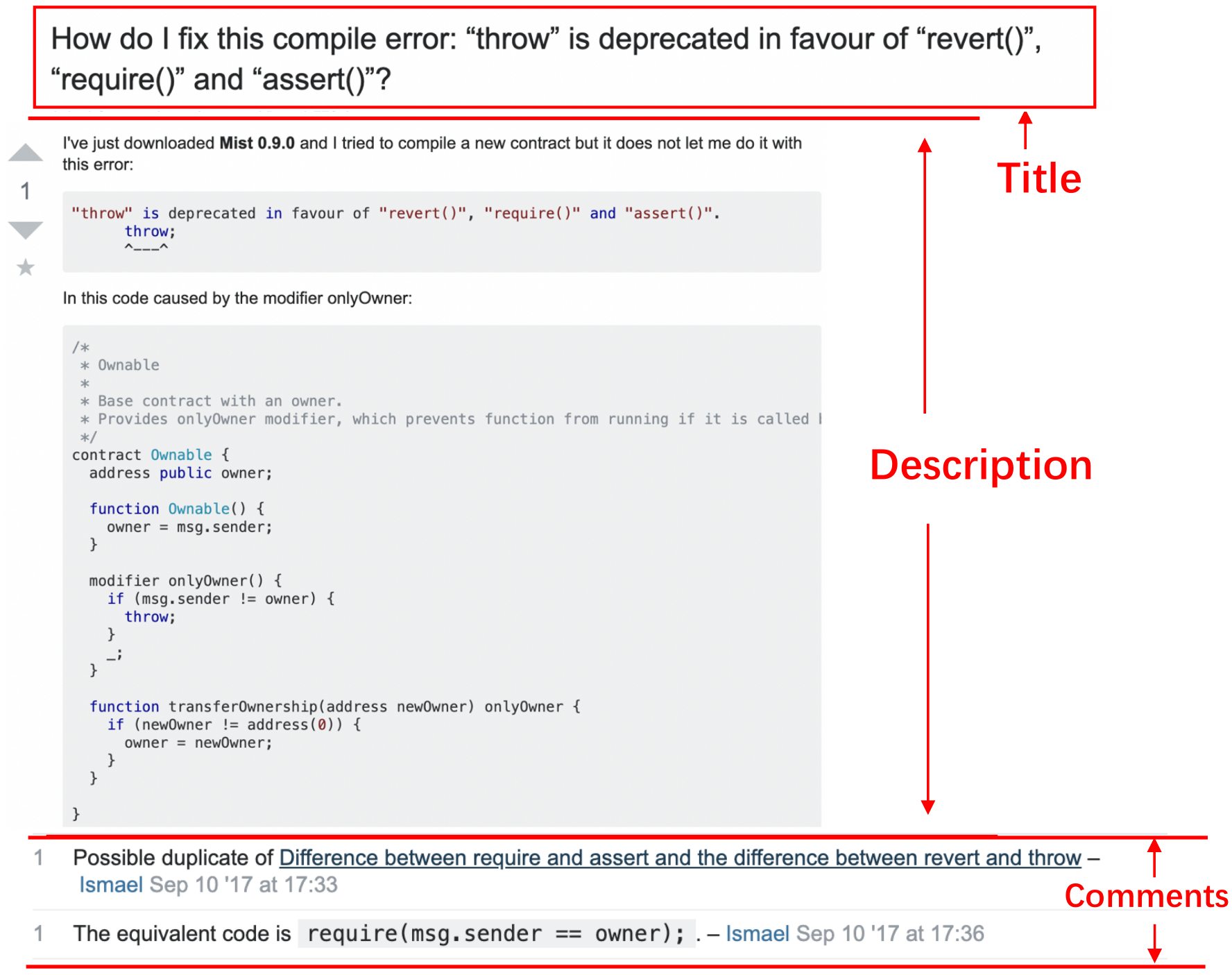} 
		\caption {Example of a Card} 
		\label{Fig:card}
	\end{center}
\end{figure} 

\textit{\underline{Iteration 1: }} We randomly chose 20\% of the cards. The same two authors first read the title and description of the card to understand the defects that the post discussed. Then, they read the comments to understand how to solve the defects. After that, they discussed the root cause of the defect. If the root cause of the card were unclear, we omitted it from our card sort. All of the themes are generated during the sorting. After this iteration, the first five categories shown in Table~\ref{tab:classification} are found. 

 \textbf{Example of categorizing a card: } Fig.~\ref{Fig:card} is an example of a card for a defect reporting post. The card contains three parts, i.e., title, description, and two comments. The two authors first read the title and description of the card to understand the contract defect(s) described by the posts. After that, they read the comments. The first comment gives a link to a previous similar post, and the second comment introduces ideas on how to fix this particular error. From the link, we can determine that the root cause of the error is because ``\textit{throw}” is deprecated since Solidity version 0.4.5. Therefore, the defect category for this card is ``\textit{Version Gaps}”.

\textit{\underline{Iteration 2: }}  Two authors independently categorized the remaining 80\% of the cards into the initial classification scheme by following the same method, described in iteration 1. During the categorizing process, they found another category named \textit{``Inappropriate Standard"}, which is common in the remaining cards. After that, they compared their results and discussed any differences. Finally, they categorized the defects into 6 themes; the detailed information is shown in Table ~\ref{tab:classification}. We used Cohen's Kappa~\cite{kappa} to measure the agreement between the two authors. Their overall Kappa value is 0.82, indicating ah strong agreement.

\noindent \textit{\textbf{3.2.5. Defining Contract Defects From Posts}}: After categorizing the filtered posts, we summarized 6 high-level root causes from \textit{StackExchange} posts. Then, the same two authors read the cards again, with the aim to find more detail behaviors for the definition of the contract defects. Finally, we summarized 16 contract defects. Following are two examples:

 \textbf{Example 1. Deprecated APIs:}  The error described in the Fig.~\ref{Fig:card} is classified into ``Version Gaps", which shows the high-level root cause. It is not difficult to find the reason of the error as the user has made use of a deprecated API, i.e., \textit{throw}. We thus conclude that we obtain a contract defect category named ``Deprecated APIs". 

\begin{figure}
	\begin{center}
		\includegraphics[width=0.49\textwidth]{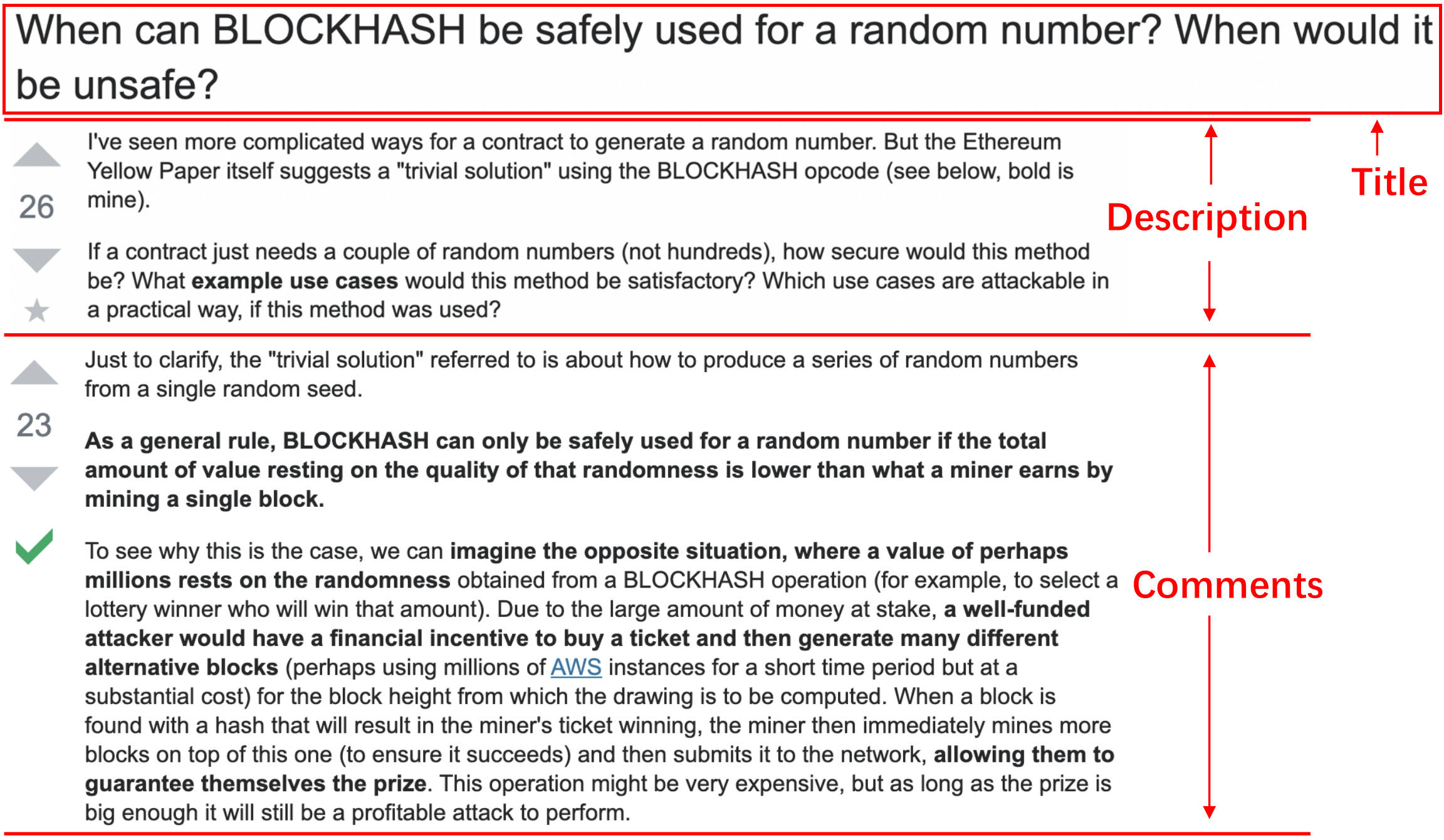} 
		\caption {Example of a Card for "Block Info Dependency"} 
		\label{Fig:card2}
	\end{center}
\end{figure} 
 \textbf{Example 2. Block Info Dependency:} Fig.~\ref{Fig:card2} is another example that belongs to the defect category ``Ethereum Features". From the post, we can determine that if the profit of the controlling contract is higher than what a miner earns by mining a single block (5 ETH), there is a high probability that the contracts will be controlled by the miner. Therefore, using \textbf{BLOCKHASH} to generate random numbers is not safe. Finally, we infer a contract defect named ``Block Info Dependency" from this card.

\noindent \textit{\textbf{3.2.6. Dataset Labeling}}: In order to assist future studies on smart contract analysis and testing, we manually identified whether the defined contract defects exist in our dataset, which consists of 578 real-world smart contracts. To build this dataset,  we first crawled all 17,013 verified smart contracts from Etherscan. Then, for the scalability reasons, we randomly chose 600 smart contracts from these 17,013 contracts. We filtered out 13 smart contracts as they do not contain any functions in their contracts. Finally, we obtained 587 smart contracts with 231,098 lines of code. The total amount of Ethers in these accounts are more than 4 million Ethers. 

\noindent \textit{\textbf{3.2.7. Defining Contract Defects From Code}}: During the process of labeling, we found some smart contracts have high similarity but also have some small differences. For example, there are two functions; the only differences for these two functions is the first function denotes a return value but does not return anything. The second function denotes the return value and correctly returns the statement. Therefore, from the difference, we defined a contract defect named ``\textit{Missing Return Statement}". We totally defined 4 contract defects from real-world smart contracts. i.e., \textit{Missing Return Statement, Strict Balance Equality, Missing Reminder}, and \textit{Greedy Contract} . Finally, we defined 20 contract defects.







\begin{table*}
	\scriptsize
	\caption{Definitions of the 20 contract defects.}
	\label{tab:difinition}
	\centering
	\begin{tabular}{p{75pt} | p{160pt} || p{75pt} | p{160pt} }
		\hline
		\textbf{Contract Defect} & \textbf{Definition} & \textbf{Contract Defect} & \textbf{Definition}\\
		\hline
		\textit{Unchecked External Calls} & Do not check the return value of external call functions. & \textit{DoS Under External Influence}  & Throwing exceptions inside a loop which can be influenced by external users\\
		\hline
		\textit{Strict Balance Equality}   & Using strict balance quality to determine the execute logic. &  \textit{Unmatched Type Assignment}& Assigning unmatched type to a value, which can lead to integer overflow  \\
		\hline
		\textit{Transaction State Dependency}  & Using tx.origin to check the permission. &  \textit{Re-entrancy} & The re-entrancy bugs. \\
		\hline
		\textit{Hard Code Address} & Using hard code address inside smart contracts. &  \textit{Block Info Dependency} & Using block information related APIs to determine the execute logic. \\
		\hline
		\textit{Nested Call}  & Executing CALL instruction inside an unlimited-length loop. &  \textit{Deprecated APIs}& Using discarded or unrecommended AIPs or instructions. \\
		\hline
		\textit{Unspecified Compiler Version}  & Do not fix the smart contract to a specific version.  &  \textit{Misleading Data Location} & Do not clarify the reference types of local variables of \textit{struct, array or mapping}.\\
		\hline
		\textit{Unused Statement } & Creating values which never be used.  &  \textit{Unmatched ERC-20 standard }& Do not follow the ERC-20 standard for ICO contracts.\\
		\hline
		\textit{Missing Return Statement}  & A function denote the type of return values but do not return anything.  &  \textit{Missing Interrupter }& Missing backdoor mechanism in order to handle emergencies. \\		
		\hline
		\textit{Missing Reminder} & Missing events to notify caller whether some functions are successfully executed. & \textit{Greedy Contract} & A contract can receive Ethers but can not withdraw Ethers.  \\		
		\hline
		\textit{High Gas Consumption Function Type} & Using inappropriate function type which can increase gas consumption. & \textit{High Gas Consumption Data Type}  & Using inappropriate data type which can increase gas consumption. \\
		\hline
	\end{tabular}	
\end{table*}

\subsection{Results}
In this part, we define and give examples of each defects. We divide these defects to five categories according to their consequences, i.e., \textit{Security defects,  Performance defects, Availability defects, Maintainability defects,} and \textit{Reusability defects}.  We first give a brief definition of each contract defects in Table \ref{tab:difinition}. Then, we give detailed definitions and code examples in the followed paragraphs:

\subsubsection{\textbf{Security Defects}}
In this subsection, we define 9 contract defects that can lead to security issues. These may be exploited by attackers to gain financial benefits or attack vulnerable contracts.

\textbf{(1) Unchecked External Calls:}
To transfer Ethers or call functions of other smart contracts, \emph{Solidity} provides a series of external call functions for raw addresses, i.e., \emph{address.send(), address.call(), address.delegatecall()}~\cite{Solidity}. Unfortunately, these methods may fail due to network errors or out-of-gas error, e.g., the 2300 gas limitation of fallback function introduced in Section \ref{background}. When errors happen, these methods will return a \emph{boolean} value (\emph{False}), but never throw an exception. If callers do not check return values of external calls, they cannot ensure whether code logic is correct.

\textbf{Example:} An example of this defect is given in \emph{Listing 1}.  In function \emph{getWinner} (L23), the contract does not check the return value of \emph{send} (L26), but the array \emph{participants} is emptied by assigning \emph{participatorID} to 0 (L25). In this case, if the \emph{send} method failed, the winner will lose 8 Ethers.

\textbf{Possible Solution:} Using \textit{address.transfer()} to instead \textit{address.send()} and \textit{address.call.value()} if possible, or Checking the return value of \textit{send} and \textit{call}. 
 
\textbf{(2) DoS Under External Influence:}
When an exception is detected, the smart contract will rollback the transaction. However, throwing exceptions inside a loop is dangerous.

\textbf{Example:}  In line 33 of \emph{Listing 1}, the contract uses \emph{transfer} to send Ethers. However,  In Solidity, \emph{transfer} and \emph{send} will limit the gas of fallback function in callee contracts to 2,300 gas~\cite{Solidity}. This gas is not enough to write to \textit{storage}, call functions or send Ethers. If one of \emph{member[i]} is an attacker's smart contract and the \emph{transfer} function (L33) can trigger an out-of-gas exception due to the 2,300 gas limitation. Then, the contract state will rollback. Since the code cannot be modified, the contract can not remove the attacker from \emph{members} list, which means that if the attacker does not stop attacking, no one can get bonus anymore.

\textbf{Possible Solution:} Avoid throwing exceptions in the body of a loop. We can return a boolean value instead of throwing an exception. For example, using ``\textit{if(msg.send(...) == false) break;}" instead of using ``\textit{msg.transfer(...)}".

\textbf{(3) Strict Balance Equality:}
Attackers can send Ethers to any contracts forcibly by utilizing \emph{selfdestruct(victim\_address)} API~\cite{Solidity}. This way will not trigger the fallback function, meaning the victim contract cannot reject the Ethers. Therefore, the logic of equal balance check will fail to work due to the unexpected ethers send by attackers.

\textbf{Example:} Attackers can send 1 Wei (1 Ether = $10^{18}$ Wei) to \emph{Contract Gamble in Listing 1} by utilizing \emph{selfdestruct} method. This method will not trigger fallback function (L13). Thus, the Ethers will not be thrown by \emph{ReceiveEth} (L16). If this attack happens, the \emph{getWinner()} (L23) would never be executed, because the \emph{getWinner} can only be executed when the balance of the contract is strictly equal to 10 Ethers (L21).

\textbf{Possible Solution:} Since the attackers can only add the amount of the balance, we can use a range to replace ``==". In this case, attackers cannot affect the logic of the programs. Using the defect in Listing 1 as an example, we can modify the code in L21 to ``\textit{if (this.balance $\geq$ 10 ether\&\& this.balance \textless 11 ether)}"

\textbf{(4) Unmatched Type Assignment:}
Solidity supports different types of integers (e.g., \textit{uint8}, \textit{uint256}). The default type of integer is \textit{uint256} which supports a range from 0 to 2 $\hat{ }$ 256. \emph{uint8} takes less memory, but only supports numbers from 0 to 2 $\hat{ }$ 8. Solidity will not throw an exception when a value exceeds its maximum value. The progressive increase is a common operation in programming, and performing an increment operation without checking the maximum value may lead to overflow.

\textbf{Example:}  The variable \emph{i} in line 30 of \emph{Listing 1} is assigned to \textit{uint8}, because 0 is in range of uint8 (0-255). If the \emph{members.length} is larger than 255, the value of \emph{i} after 255 is 0. Thus, the loop will not stop until running out of gas or balance of account is less than 0.1.

\textbf{Possible Solution:} Using \textit{uint} or \textit{uint256} if we are not sure of the maximum number of loop iterations.  

\textbf{(5) Transaction State Dependency:}
Contracts need to check whether the caller has permissions in some functions like \emph{suicide} (L33 in \emph{Listing 1}). The failure of permission checks can cause serious consequences. For example, if someone passes the permission check of \emph{suicide} function, he/she can destroy the contract and stole all the Ethers. \emph{tx.origin} can get the original address that kicked off the transaction, but this method is not reliable since the address returned by this method depends on the transaction state.

\textbf{Example:}  We can find this defect in line 8 of \emph{Listing 1}. The contract uses \emph{tx.origin} to check whether the caller has permission to execute function \emph{suicide} (L35). However, if an attacker uses function \emph{attack} in \emph{Listing 4} to call \emph{suicide} function (L35 in \emph{Listing 1}), the permission check will fail. \emph{suicide} function will check whether the sender has permission to execute this function. However, the address obtained by \emph{tx.origin} is always the address who creates this contract (0xdCad...d1D3AD \emph{L12 in Listing 1}). Therefore, anyone can execute the \emph{suicide} function and withdraw all of the Ethers in the contract.

\textbf{Possible Solution:} Using \emph{msg.sender} to check the permission instead of using \emph{tx.orign}.

\textbf{(6) Block Info Dependency:}
Ethereum provides a set of APIs (e.g.,  block.blockhash, block.timestamp) to help smart contracts obtain block related information, like timestamps or hash number. Many contracts use these pieces of block information to execute some operations. However, the miner can influence block information; for example, miners can vary block time stamp by roughly 900 seconds~\cite{Block_validation_algorithm}. In other words, block info dependency operation can be controlled by miners to some extent.

\textbf{Example:}  In Listing 1 line 25, the contract uses blockhash to generate which member is the winner. However, the gamble is not fair because miners can manipulate this operation. 

\textbf{Possible Solution:} To generate a safe random number in Solidity, we should ensure the random number cannot be controlled by a single person, e.g., a miner. We can use the information of users like their addresses as their input numbers, as their distributions are completely random. Also, to avoid attacks, we need to hide the values we used from other players. Since we cannot hide the address of users and their submitted values, a possible solution to generate a random number without using block related APIs is using a hash number. The algorithm has three rounds: 

\textit{Round 1}: Users obtain a random number and generate a hash value in their local machine. The hash value can be obtained by \textit{keccak256}, which is provided by Solidity. After obtaining the random number, users submit the hash number.

\textit{Round 2}: After all users submit their hash number, users are required to submit their original random number. The contract checks whether the original number can generate the same hash number.

\textit{Round 3}: If all users submit the correct original numbers, the contract can use the original numbers to generate a random number. 

\textbf{(7) Re-entrancy:}
Concurrency is an important feature of traditional software. However, Solidity does not support it, and the functions of a smart contract can be interrupted while running. Solidity allows parallel external invocations using \textit{call} method. If the callee contract does not correctly manage the global state, the callee contract will be attacked -- called a re-entrancy attack.

\textbf{Example:}  \emph{Listing 2} shows an example of re-entrancy. The \textit{Attacker} contract invokes \textit{Victim} contract's withDraw() function in Line 11. However, \textit{Victim} contract sends Ethers to \textit{attacker} contract (L6) before resetting the balance (L7). Line 6 will invoke the fallback function (L9) of \textit{attacker} contract and lead to repeated invocation.

\textbf{Possible Solution:} Using \textit{send()} or \textit{transfer} to transfer Ethers.  \textit{send()} and \textit{transfer}  have \textit{gas limitation} of  2300 if the recipient is a contract account, which are not enough to transfer Ethers. Therefore, these two functions will not cause Re-entrancy. 

\begin{lstlisting}[caption={Attacker contract can attack Victim contract by utilizing Re-entrancy }]
contract Victim {
  mapping(address => uint) public userBalannce;
  function withDraw(){
	uint amount = userBalannce[msg.sender];
	if(amount > 0){
	  msg.sender.call.value(amount)();
	  userBalannce[msg.sender] = 0;}} ...}
contract Attacker{
  function() payable{
	Victim(msg.sender).withDraw();}
  function reentrancy(address addr){
	Victim(addr).withDraw();} ...}
\end{lstlisting}

\textbf{(8) Nested Call:}
Instruction \emph{CALL} is very expensive (9000 gas paid for a non-zero value transfer as part of the CALL operation~\cite{yellowpaper}). If a loop body contains CALL operation but does not limit the number of times the loop is executed, the total gas cost would have a high probability of exceeding the gas limitation because the number of iterations may be high and it is hard to know its upper limit.

\textbf{Example:}  In \emph{Listing 1}, the function \emph{giveBonus} (line 28) uses \emph{transfer} (L33) which generates \emph{CALL} to send Ethers. Since the \emph{members.length} (L30) does not limit its size, \emph{giveBonus} has a probability to cause out of gas error. When this error happens, this function can not be called anymore because there is no way to reduce the \emph{members.length}.

\textbf{Possible Solution:} The developers should estimate the maximum number of loop iterations that can be supported by the contract and limit these loop iterations. 

\textbf{(9) Misleading Data Location:}
In traditional programming languages like \emph{Java} or \emph{C}, variables created inside a  function are local variables. Data is stored in \textit{memory} and the \textit{memory} will be released after the function exits. In Solidity, the data of \emph{struct, mapping, arrays} are stored in \textit{storage} even they are created inside a function. However, since \textit{storage} in solidity is not dynamically allocated, storage variables created inside a function will point to the \textit{storage slot}\footnote{Each storage variables has its own storage slot to identify its position.} 0 by default~\cite{Solidity}. This can cause unpredictable bugs.

\textbf{Example:}  Function \emph{reAssignArray} (L6) in \emph{Listing 3} creates a local variable \emph{tmp}. The default data location of \emph{tmp} is \textbf{storage}, but EVM cannot allocate storage dynamically. There is no space for \emph{tmp}, but instead, it will point to the storage slot \emph{0 (variable} in L3 of \emph{Listing 3)}. For the result, once function \emph{reAssignArray} is called, the variable \emph{variable} will add 1, which can cause bugs for the contract.

\textbf{Possible Solution:} Clarifying the data location of \emph{struct, mapping}, and \emph{arrays} if they are created inside a function. 

\begin{lstlisting}[caption={DefectExample}]
pragma solidity ^0.4.25;/*Unspecified Compiler Version*/
contract DefectExample{
  uint variable;
  uint[] investList;
  function() payable{}
  function reAssignArray(){
	/*Misleading Data Location*/
	uint[] tmp;
	tmp.push(0);
	investList = tmp;}
  function changeVariable(uint value1, uint value2){
	/*Unused Statement*/
	uint newValue = value1;
	variable = value2;}
	/*High Gas Consumption Function Type*/
  function highGas(uint[20] a) public returns (uint){
	return a[10]*2;}
  function lowGas(uint[20] a) external returns (uint){
	return a[10]*2;}}
\end{lstlisting}

\begin{lstlisting}[caption={An attacker contract by utilizing Transaction State Dependency. }]
contract attacker{
	...
	function attack(address addr, address myAddr){
		Gamble gamble = Gamble(addr);
		gamble.suicide(myAddr);}}
\end{lstlisting} 

\subsubsection{\textbf{Availability Defects}} 
We define 4 contract defects related to availability. These may not be utilized by attackers but are bad designs for contracts that can lead to potential errors or financial loss for the caller. 

\textbf{(1) Unmatched ERC-20 Standard:}
ERC-20 Token Standard~\cite{erc20} is a technical standard on Ethereum for implementing tokens of cryptocurrencies. It defines a standard list of rules for Ethereum tokens to follow within the larger Ethereum ecosystem, allowing developers to predict the interaction between tokens accurately. These rules include how the tokens are transferred between addresses and how data within each token is accessed. The function name, parameter types and return value should strictly follow the ERC20 standard. ERC-20 defines 9 different functions and 2 events to ensure the tokens based on ERC20 can easily be exchanged with other ERC20 tokens. However, we find that many smart contracts miss return values or miss some functions.

\textbf{Example:} \emph{transfer} and \emph{transferFrom} are two functions defined by ERC20. They are used to transfer tokens from one account to another. ERC20 defines that these two functions have to return a \emph{boolean} value, but many smart contracts miss this return value, leading to errors when transferring tokens.

\textbf{Possible Solution:} Checking that the contract has strictly followed the ERC20 standard.

\textbf{(2) Missing Reminder:}
Other programs can call smart contracts through the contracts' \textit{Application Binary Interface (ABI)}. ABI is the standard way to interact with contracts in the Ethereum ecosystem, both from outside the blockchain and for contract-to-contract interaction.  However, the ABIs can only tell the caller what the inputs and outputs of a function are, but it will not inform them whether the function call is successful or not. Throwing an event to notify a caller whether the function is successfully executed can reduce unnecessary errors and gas waste.

\textbf{Example:}  A typical scenario of this contract defect is missing reminders when receiving Ethers. In \emph{Listing 1}, users may not understand the game rules clearly, and send Ethers which not equal to 1 Ether (line 16-17). However, the smart contract will check whether the received Ether is equal to 1 Ether, then the Ether will return back. There are several reasons for invoking failures.  For example, the user may mistakenly believe the error is caused by network and resend the Ethers, which can lead to gas waste. Adding reminders (throwing events) to notify caller whether some functions are successfully executed can avoid unnecessary failure.

\textbf{Possible Solution:} Adding reminders for functions that are interacting with the outside. 

\textbf{(3) Missing Return Statement:}
Some functions denote return values but do not return anything. For these, EVM will add a default return value when compiling the code to bytecode. Since the callers may not know the source code of the callee contract, they may use the return value to handle code execution and lead to unpredictable bugs.

\textbf{Example:} Function \emph{giveBonus} (L28) in \emph{Listing 1} declares the return type \emph{bool}, but the function does not return \emph{true} or \emph{false}. Then, EVM will assign the default return value as \emph{false}. If developers call this function, the return value will always be the \emph{false} and some functions in the caller contracts may never be executed.

\textbf{Possible Solution:} Adding the return statements for each function. 

\textbf{(4) Greedy Contract:}
A contract can withdraw Ethers by sending Ethers to another address or using \emph{selfdesturct} function. Without these withdraw-related functions, Ethers in contracts can never be withdrawn and will be locked forever.  We define a contract to be a greedy contract if the contract can receive ethers (contains payable fallback function) but there is no way to withdraw them.

\textbf{Example:} In \emph{Listing 3}, the contract has a \emph{payable fallback function} in line 5, which means this contracts can receive Ethers. However, the contracts cannot send Ethers to other contracts or addresses. Therefore, the Ethers in this contract will be locked forever.

\textbf{Possible Solution:} Adding withdraw method if the contract can receive Ethers. 


\subsubsection{\textbf{Performance Defects}} 
We define 3 contract defects related to performance. The contracts with these defects can increase their gas cost. 

\textbf{(1) Unused Statement: }
If function parameters or local variables do not affect any contract statements nor return a value, it is better to remove these to improve code readability.

\textbf{Example:} function parameter \emph{value1} and local variable \emph{newValue} in function \emph{changeVariable} (L11 of \emph{Listing 3}) are useless, because they never affect contract statements nor return values. Although the compiler will remove these useless statements when compiling source code to binary code, these can reduce contract readability.

\textbf{Possible Solution:}  Removing all unused statements in the contract to make it easier to read. 

\textbf{(2) High Gas Consumption Function Type:}
For \emph{public} functions, Solidity immediately copies function arguments (\emph{Arrays}) to \textit{memory}, while \emph{external} functions can read directly from \emph{calldata}~\cite{yellowpaper}. Memory allocation is expensive, whereas reading from \emph{calldata} is cheap. To lower gas consumption, if there are no internal functions call this function and the function parameters contain \emph{array}, it is recommended to use \emph{external} instead of \emph{public}.

\textbf{Example:}  In \emph{Listing 3}, function \emph{highGas} (L16) and function \emph{lowGas} (L18) have the same capabilities. The only difference is that \emph{highGas} is modified by \emph{public} which can be called by external and internal functions. \emph{lowGas} is modified by \emph{external} which can only be called by external. Calling function \emph{highGas} costs 496 gas while calling \emph{lowGas} only costs 261 gas.

\textbf{Possible Solution:} Using \emph{external}  instead of \emph{public} if the function can only be called by external.  

\textbf{(3) High Gas Consumption Data Type:}
\emph{bytes} is dynamically-sized byte array in Solidity, \emph{byte[]} is similar with \emph{bytes}, but \emph{bytes} cost less gas than \emph{byte[]} because it is packed tightly in \emph{calldata}. EVM operates on 32 bytes a time, \emph{byte[]} always occupy multiples of 32 bytes which means great space is wasted but not for \emph{bytes}. Therefore, \emph{bytes} takes less storage and costs less gas. To lower gas consumption, it is recommended to use \emph{bytes} instead of \emph{byte[]}.

\textbf{Example:} Replacing \emph{byte[]} by \emph{bytes} can save a small amount of gas for each function call. However, as the contract is called more times, a large amount of gas can potentially be saved.

\textbf{Possible Solution:} Using \emph{bytes} instead of byte[]. 

\subsubsection{\textbf{Maintainability Defects}} 
We define 2 contract defects related to maintainability. These contract defects can shorten the life cycle of the contract. 

\textbf{(1) Hard Coded Address:}
Since we cannot modify smart contracts after deploying them, hard coded addresses can lead to vulnerabilities.

\textbf{Example:}  There are two main kinds of errors this contract defect can lead to. The first is \textit{Illegal Address}. Ethereum uses a mixed-case address checksum to verify whether an address is legal or not. The rule is defined in EIP-55~\cite{EIP55}. There is an error address in line 12 of Listing 1. The owner address is an illegal address, the last bit of the address should be `F', but by mistake, it becomes `D'. The illegal address makes no one that can withdraw the amount of this contract. The second is \textit{Suicide Address}. \textit{selfdestruct} function (L36) can remove the code from the blockchain and make the contract become a suicide contract, but it is potentially dangerous. If someone sends Ether to suicide contracts, the Ether will forever be lost. \textit{receiver} (L38) is a smart contract who contains \textit{selfdestruct} function. Its address is hardcoded in line 38 of Listing 1 and cannot be modified. If the \textit{receiver} performed the \textit{selfdestruct} function, it will become a suicide contract. All the Ethers sent to  \textit{receiver} will be lost forever.

\textbf{Possible Solution:} Removing the hard coded addresses and inputting the addresses as function parameters. 

\textbf{(2) Missing Interrupter:}
When bugs are detected by attackers, they can attack the contracts and steal their Ethers. The DAO lost \$50 million Ethers due to a bug in the code that allowed an attacker to draw off the Ethers~\cite{DAOAttack} repeatedly. The interrupter is a mechanism to stop the contract when bugs are detected. We cannot modify contracts after deploying them to the blockchain. However, if a contract contains interrupter, the owner of the victim contract can reduce their losses. 

\textbf{Example:} When bugs are found in \emph{Listing 1}, the Ethers on the contract can be stolen by attackers. Fortunately, the contract contains an interrupter on \textit{suicide function}  (L35). So, the owner of the contract can call \emph{suicide}. Then, the remain Ethers will be sent to the given address. After fixing the bugs, the contracts can be redeployed.

\textbf{Possible Solution:} The easiest interrupter is adding a \emph{selfdestruct} function~\cite{Solidity}, Ethers on the contracts can be withdrawn and the contracts destroyed when attacks happen. Adding an interrupter to the contracts, if the contract holds a large amount of Ethers.

\subsubsection{\textbf{Reusability Defects}} 
We define 2 contract defects related to reusability. These contract defects can increase the difficulty of code reuse. 

\textbf{(1) Deprecated APIs:}
Solidity is a young and evolving programming language. Some APIs will be discarded or updated in the future. In this case, Solidity documentation usually uses warning to inform developers that some APIs will be deprecated in the future. These APIs might still be supported by the current compiler version. However, if developers use these APIs, they might need to refactor the code for the code reuse, which leads to resource waste.

\textbf{Example:}   \emph{CALLCODE} operation will be discarded in the future~\cite{Solidity}, \emph{throw, suicide, sha3} are replaced by \emph{revert, selfdestruct, keccak256} respectively in the recent version.

\textbf{Possible Solution:} Following the latest Solidity document and using the latest APIs. 

\textbf{(2) Unspecified Compiler Version:}
Different versions of Solidity may contain different APIs/instructions. In Solidity programming, multiple APIs only be supported in some specific versions. If a contract do not specify a compiler version, developers might encounter compile errors in the future code reuse because of the version gap. 

\textbf{Example:} In the first line of \emph{Listing 3}, \emph{pragma solidity $\hat{ }$ 0.4.25 } means that this contract supports compile version 0.4.25 and above (except for v0.5.0) while \emph{pragma solidity 0.4.25 } means that the contract only supports compile version 0.4.25. Since it is hard to foresee the language constructions in the future version, it is recommended to indicate a specific compiler version to avoid unnecessary bugs.  	

\textbf{Possible Solution:} Fixing the compiler version used by the contract. 




\section{RQ2: Practitioners' Perspective}
\label{Evaluation}

\subsection{Motivation}
To validate whether our defined contract defects are harmful, we created an online survey to collect opinions from real-world smart contract developers.

\begin{table*}
	\scriptsize	
	\caption{Survey results, distributions, and  impacts of the 20 contract defects. }
	\label{tab:result}
	\centering
	\begin{tabular}{p{90pt} <{\centering} p{40pt} <{\centering}  p{15pt} <{\centering}  p{40pt} <{\centering} p{17pt} <{\centering} | p{90pt} <{\centering} p{40pt} <{\centering}  p{20pt} <{\centering}  p{40pt} <{\centering} p{15pt} <{\centering} }
		\hline
		Contract Defect & Distribution & Score & \#Defects & Impacts & Contract Defect & Distribution & Score & \#Defects & Impacts\\
		\hline
		
		Unchecked External Calls & \includegraphics[ height=3mm]{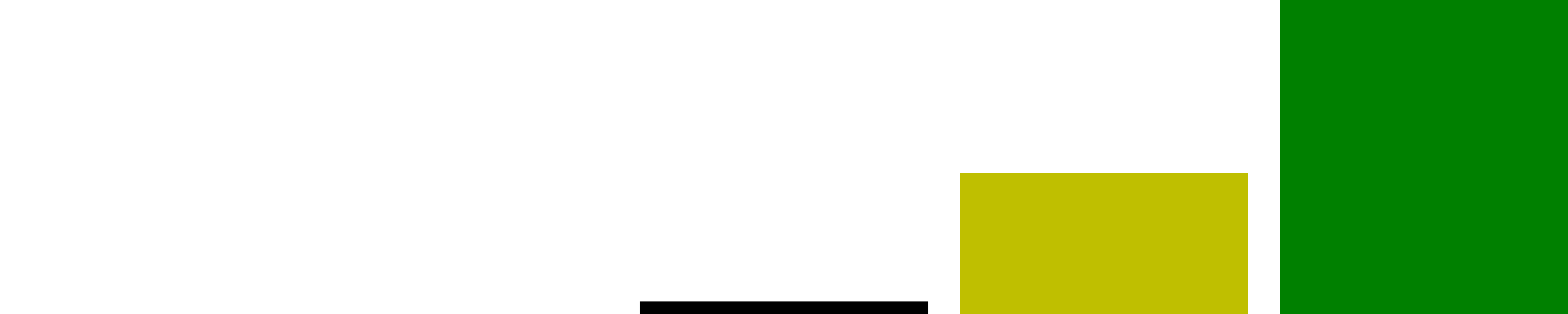} & 4.50 & 25 (4.26\%) & IP3 & DoS Under External Influence  & \includegraphics[ height=3mm]{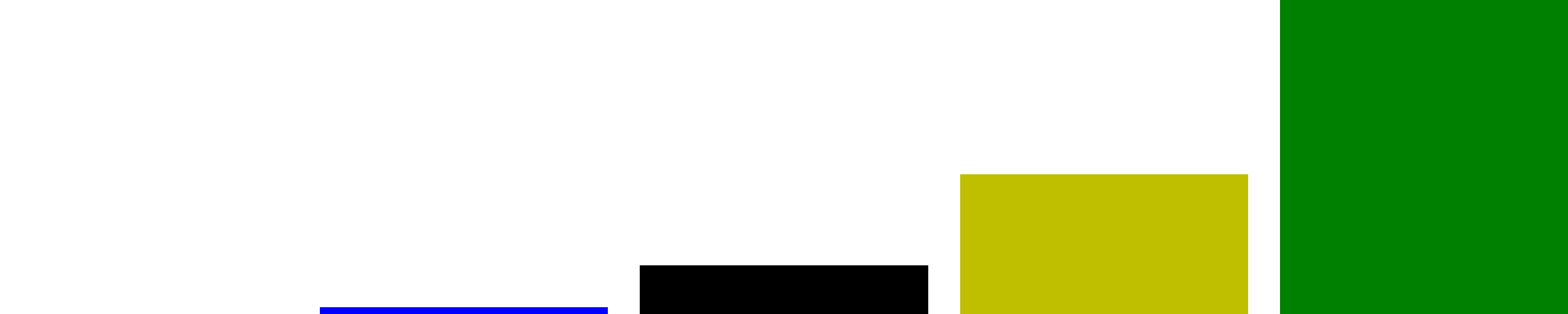} & 4.31  & 6 (1.02\%) & IP2\\
		
		Strict Balance Equality  & \includegraphics[ height=3mm]{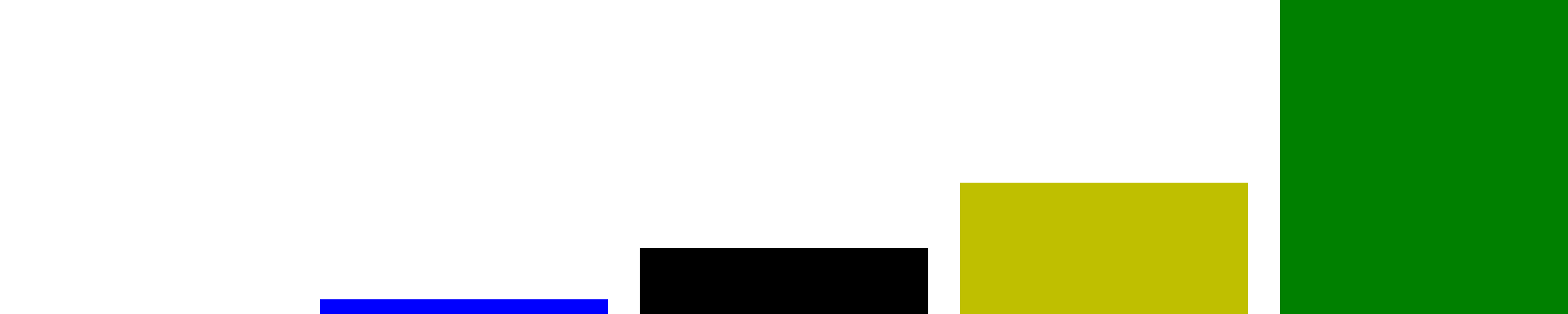}  & 4.28 & 5 (0.85\%)  & IP2 &  Unmatched Type Assignment& \includegraphics[ height=3mm]{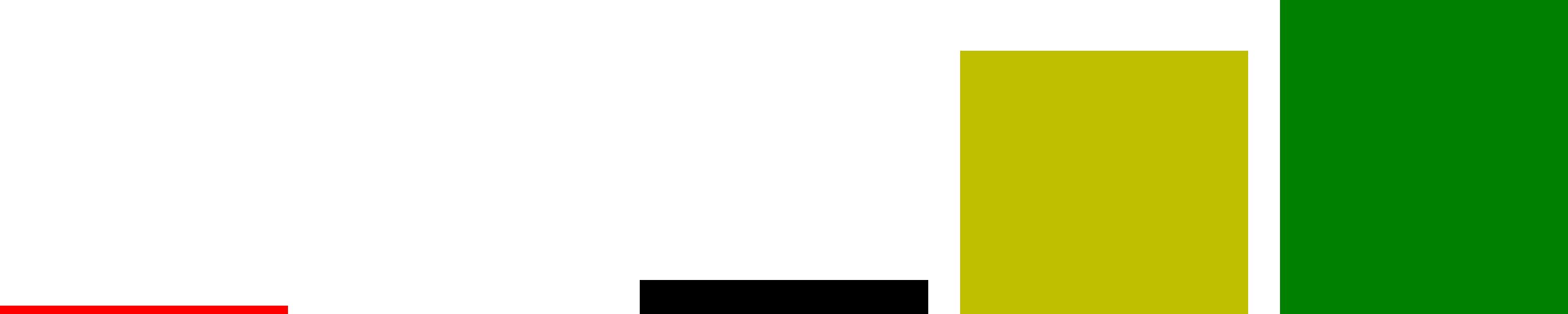} & 4.42 & 22 (3.75\%) & IP2 \\
		
		Transaction State Dependency  & \includegraphics[ height=3mm]{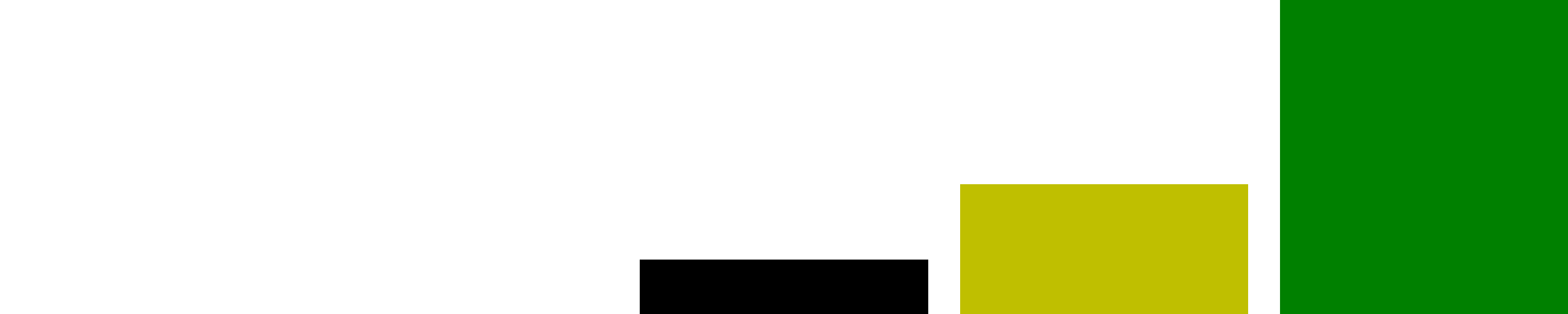} & 4.54 & 5 (0.85\%) & IP1 &  Reentrancy & \includegraphics[ height=3mm]{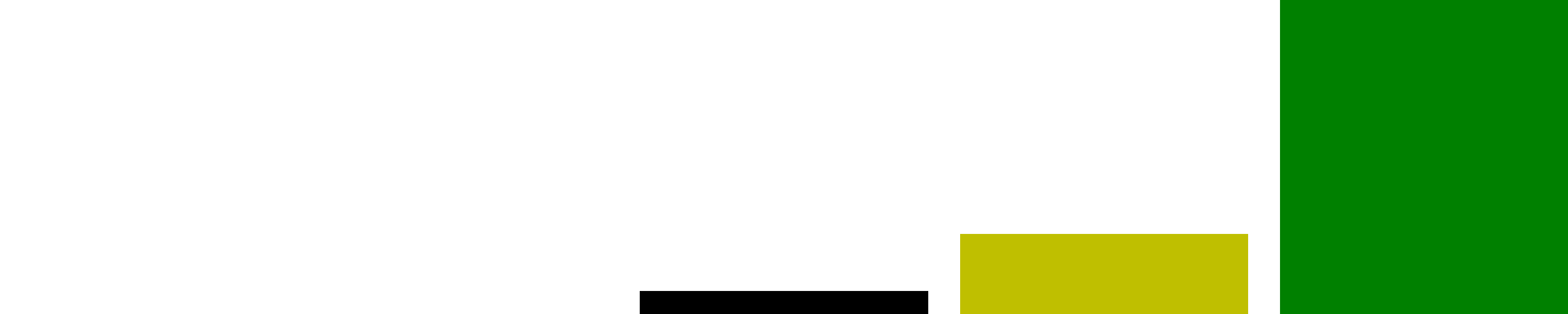} & 4.66 & 12 (2.04\%) & IP1\\
		
		Hard Code Address &  \includegraphics[ height=3mm]{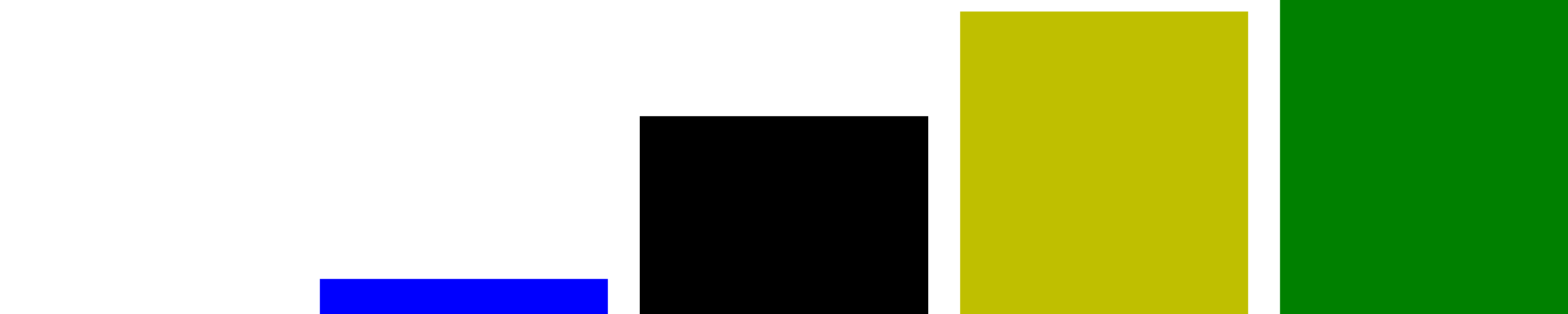}  &4.10 & 84 (14.31\%) & IP3 &  Block Info Dependency &  \includegraphics[ height=3mm]{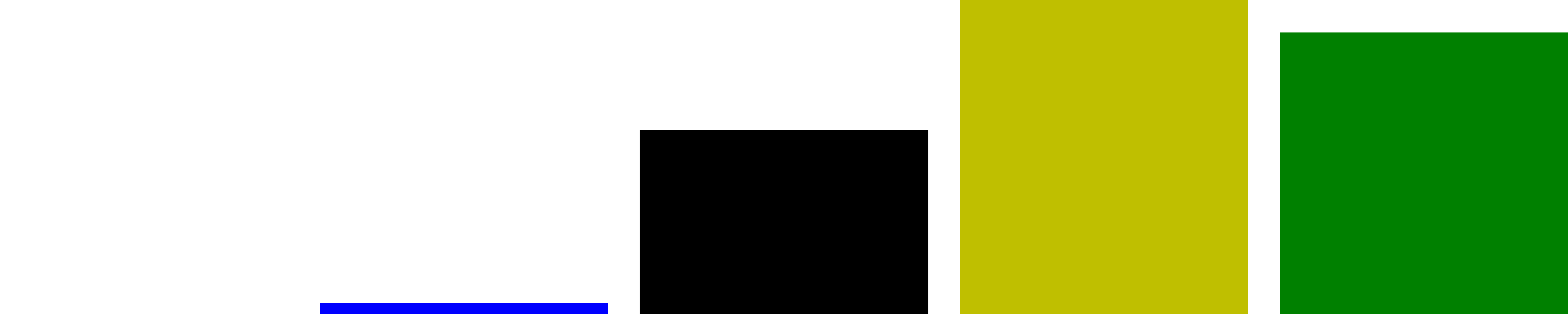} & 4.05 & 42 (7.16\%) & IP3 \\
		
		Nested Call  & \includegraphics[ height=3mm]{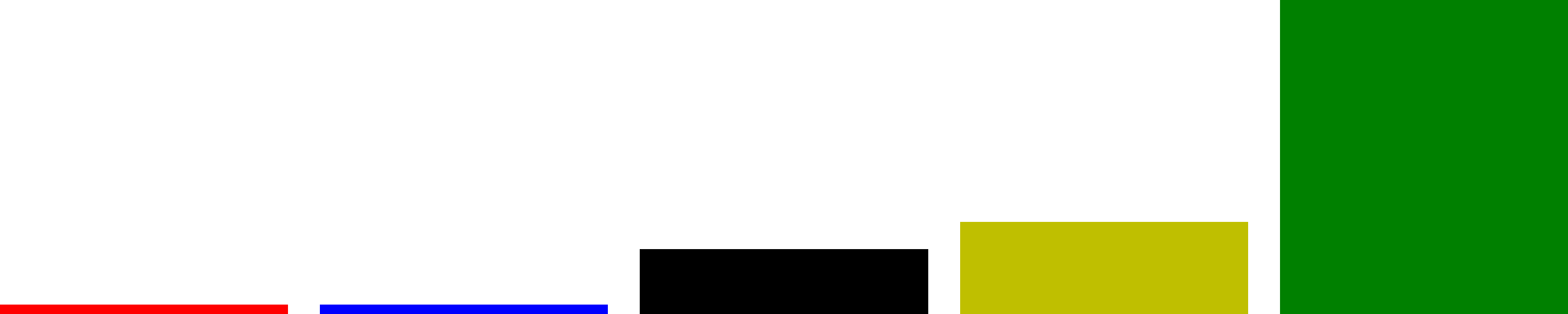} & 4.45 & 13 (2.21\%) & IP2 &  Deprecated APIs& \includegraphics[ height=3mm]{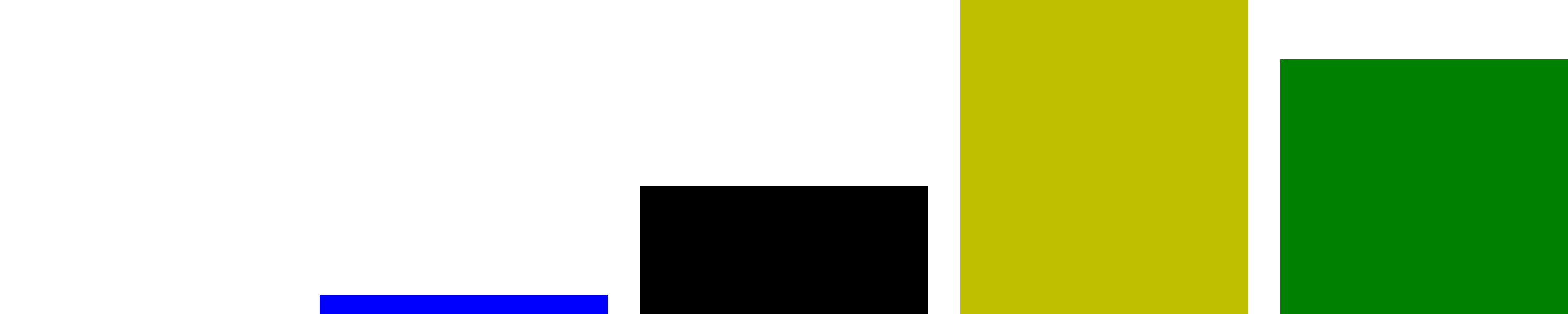} & 4.06 & 247 (42.08\%) & IP5 \\
		
		Unspecified Compiler Version  & \includegraphics[ height=3mm]{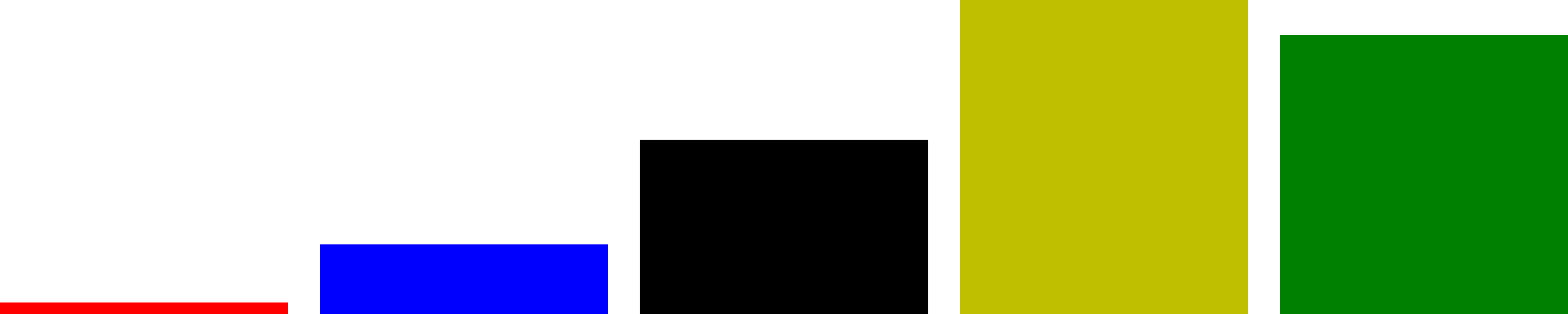} & 3.84 & 532 (90.63\%) & IP5  &  Misleading Data Location & \includegraphics[ height=3mm]{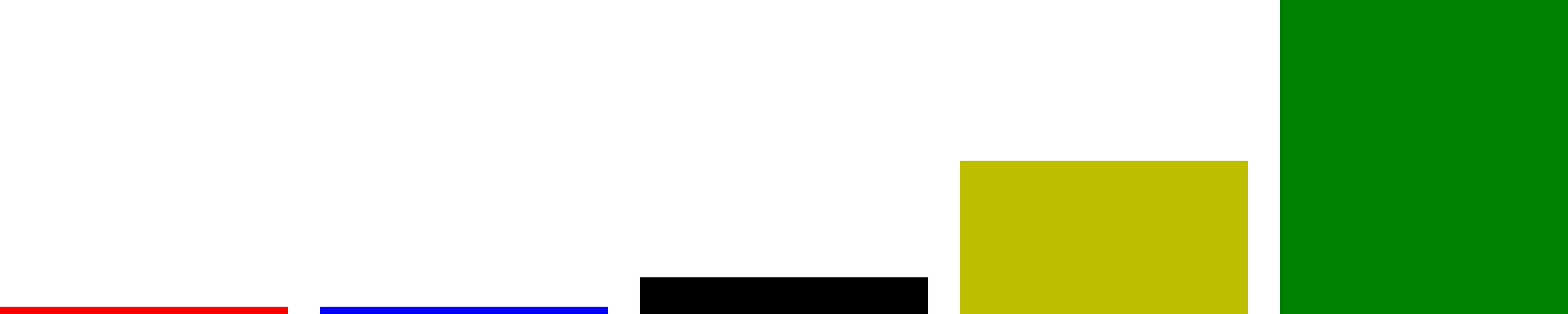} & 4.28 & 1 (0.17\%) & IP2 \\
		
		Unused Statement  & \includegraphics[ height=3mm]{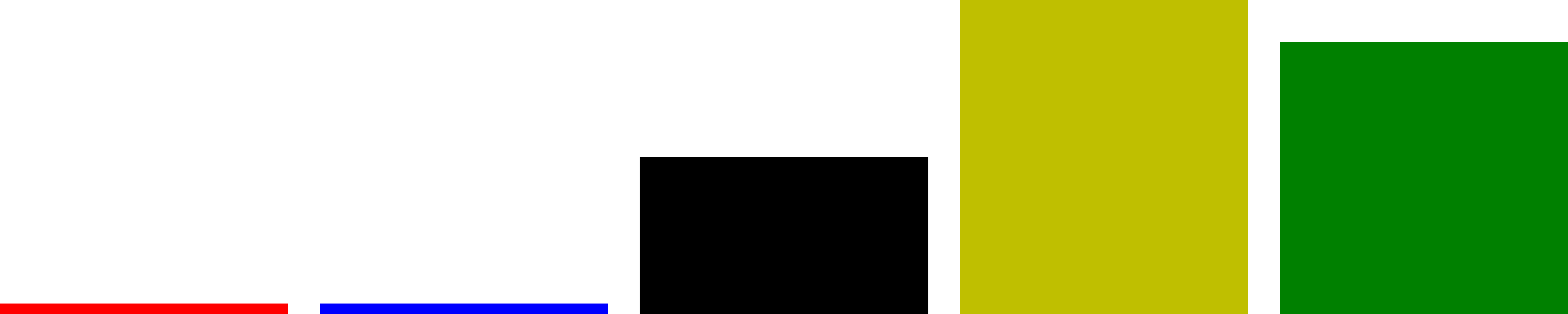} & 4.04 & 10 (1.70\%) & IP5 &   Unmatched ERC-20 standard & \includegraphics[ height=3mm]{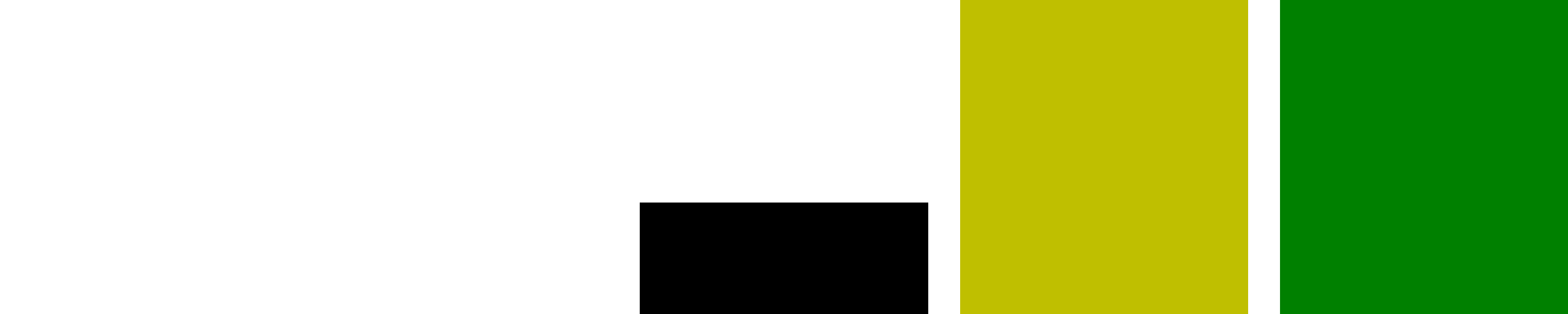} & 4.29 & 45 (7.67\%) & IP4 \\
		
		Missing Return Statement  & \includegraphics[ height=3mm]{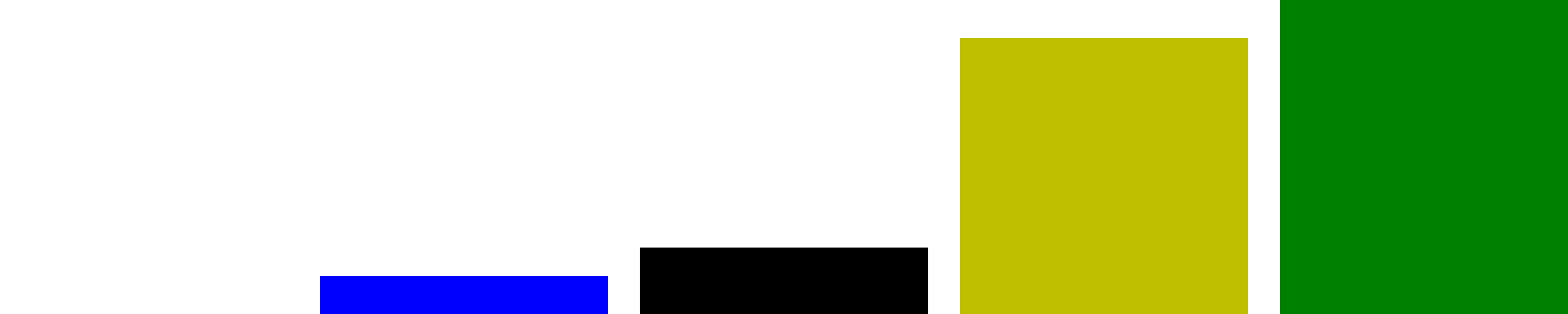} & 4.16 & 263 (44.80\%) & IP4 &  Missing Interrupter & \includegraphics[ height=3mm]{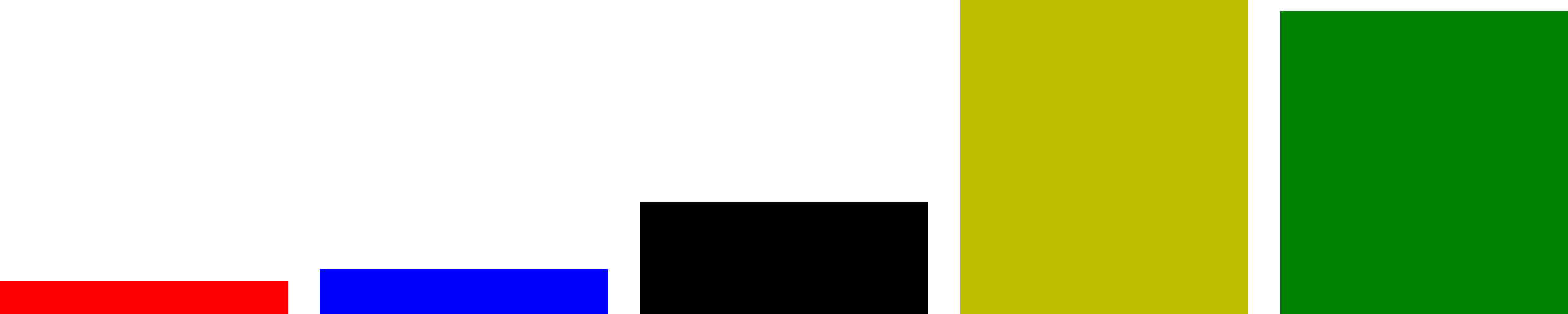} & 4.06 & 523 (89.10\%) & IP4 \\		
		
		Missing Reminder & \includegraphics[ height=3mm]{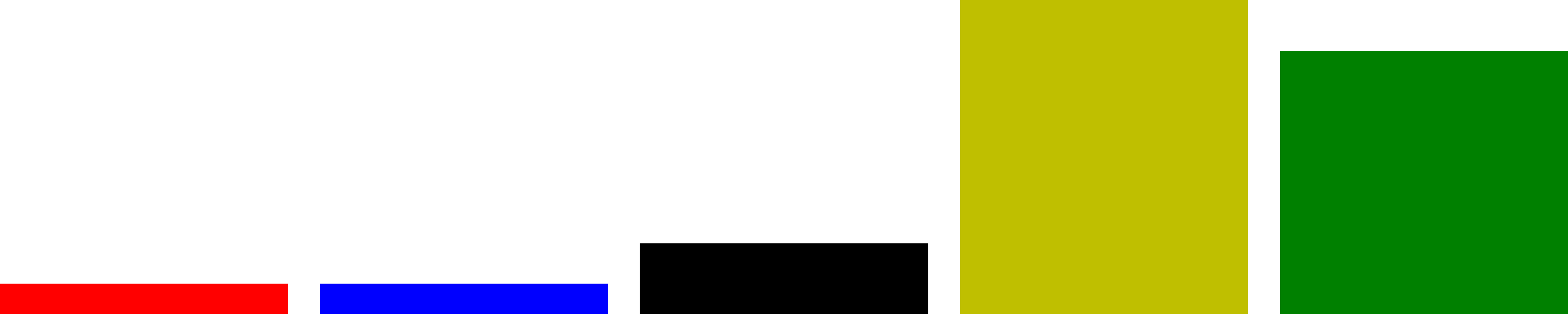} & 4.06 & 27 (4.60\%) & IP4 & Greedy Contract & \includegraphics[ height=3mm]{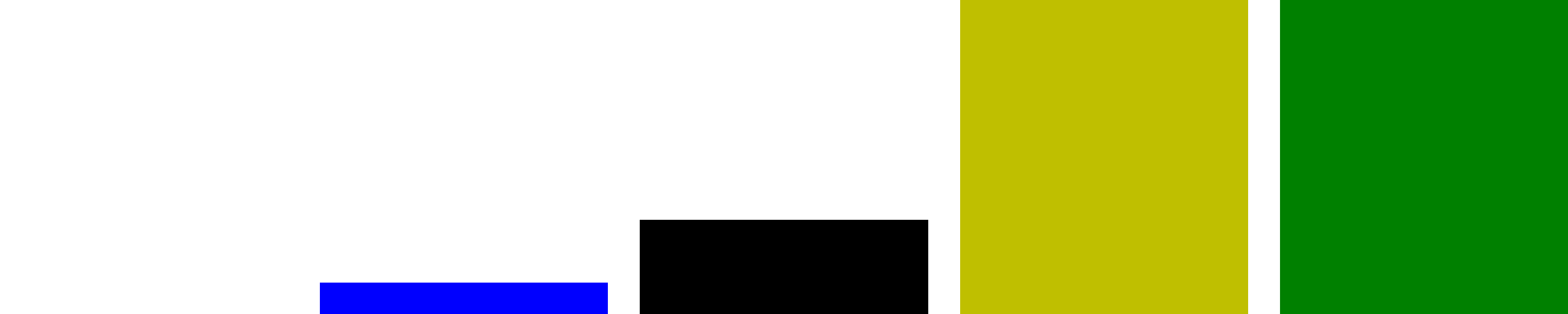} & 4.25 & 6 (1.02\%) & IP3\\		
		
		High Gas Consumption Function Type & \includegraphics[ height=3mm]{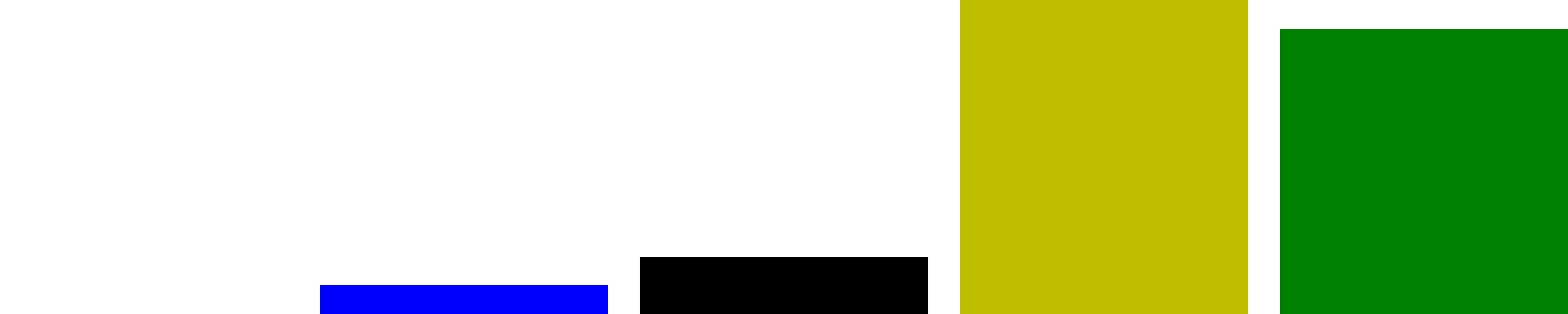}  & 4.08 & 422 (71.89\%) & IP5& High Gas Consumption Data Type  & \includegraphics[ height=3mm]{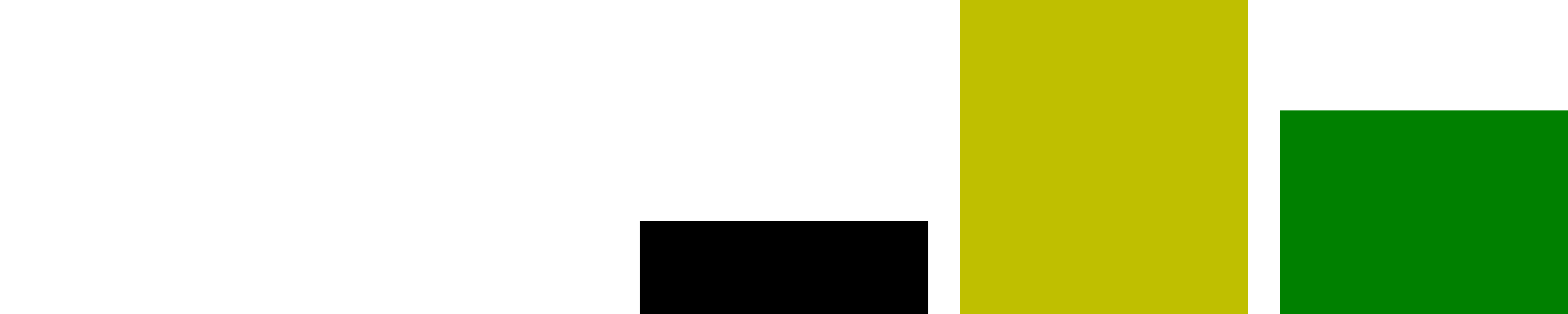} & 4.07 & 0 (0\%)  & IP5 \\
		\hline
	\end{tabular}	
\end{table*}

\subsection{Approach}
\subsubsection{Validation Survey}
We followed the instructions of Kitchenham et al. ~\cite{kitchenham2008personal} for personal opinion surveys and utilized an anonymous survey~\cite{tyagi1989effects} to increase response rates. Respondents can choose to leave an email address, as all respondents could choose to take part in a raffle to win two \$50 Amazon gift cards.  We first conducted a small scale survey to test and refine our questions. These participants give feedback about: (1) whether the expression of the contract defects is clear and easy to understand, and (2) whether the length of each question is suitable. Finally, we modified our survey based on the feedback we collected. 


\subsubsection{Survey Design}
To help respondents better understanding the aim of our survey, we explained what is contract defect at the beginning of the survey and gave detailed definitions and examples of the 20 contract defects in related questions.  We first captured the following pieces of information to collect demographic information about the respondents:

\textbf{Demographics: }
\begin{itemize}\setlength{\itemsep}{1pt}
	
	\item Professional smart contract developer? : Yes / No
	
	\item Involved in open source software development? : Yes / No
	
	\item Main role in developing smart contract.
	
	\item Experience in years
	
	\item Current country of residence
	
	\item Highest educational qualification
	
	
\end{itemize}		

\textbf{Examples of Contract Defects:}
Next, we gave detailed definitions and examples of the 20 contract defects. We asked respondents to rate the importance of these contract defects, i.e., removing them can improve the security, reliability, or usability of a project.  Since some of the defined contract defects are not easy to understand, we added an option ``\textit{I don't understand}" to ensure results are reliable.  Finally, we give each question six options (i.e., Very important,  Important, Neutral, Unimportant, Very unimportant and I don't understand). We also give each question a textbox to enable respondents to give their opinions.

\textbf{Other Questions:}  We give a textbox so respondents can tell us if they have any other comments, questions, or concerns.

\subsubsection{Recruitment of Respondents}
In order to get a sufficient number of respondents from different backgrounds, we first sent our survey to our partners who are working or study in world-famous companies or academic institutions. We sent our email to 1489 practitioners who contribute to open source smart contract related projects on GitHub. All respondents could enter their email to take part in a raffle to win two \$50 Amazon gift cards.

\subsection{Results}
We totally received 138 responses (The response rate is about 9.27\%) from 32 different countries, and we received 84 comments on our defined contract defects. 113 (81.88\%) of these respondents are involved in open source software development efforts. The top two countries in which the respondents reside are China (38.41\%) and USA (7.97\%). The average years of experience in developing smart contracts are 1.95 years. Since the Ethereum was published only in late 2015, we believe the average year of 1.95 years shows that the respondents have good experience in developing smart contracts. We do not remove the feedback from developers with little experience as their feedback is also very useful as they might be the ones actually authoring the contracts with defects. Among these respondents, 89 (64.49\%), 17 (12.32\%), 16 (11.59\%), 7 (5.07\%) described their job roles as development, testing, management and security audit respectively. The other 9 responses said they have multiple roles.

Table \ref{tab:result} shows the results of our survey. The first column indicates each contract defect and the second column illustrates the distribution of respondents' choice. The distribution is from ``Very unimportant" (left-most red bar) to ``Very important" (right-most green bar). To clearly show the result, we give each option a score and count the weighted average score which is shown in the third column. To be specific, we give ``very important" a score 5 and give ``very unimportant" a score 1.

We received very positive feedback from developers with \textbf{ almost all contract defects' scores are larger than 4, and the average score is 4.22}. The score of ``Unspecified Compiler Version" is 3.84 but it is also a positive score. To understand the reasons, we reviewed comments about this defect. We found that many developers who voted ``unimportant" mentioned the difference among different minor versions in the same major version  (e.g., 0.4.19 and 0.4.20) is small. However, they admitted that the difference among different major versions (e.g., 0.4.0 and 0.5.0) is significant. Some developers gave comments that removing this contract defect is very important when they want to reuse code in the future. Besides, we also found many examples from StackExachange posts that many developers failed to compile the contracts because these contracts do not specify their compiler versions. Therefore, we believe this defect is important on code reuse.

``Missing Interrupter”, ``Missing Reminder”, and ``Unspecified Compiler Version” received the top three most negative feedbacks (``Unimportant” and ``Very unimportant”). For \textbf{``Missing Interrupter’’}, 5 developers mentioned that adding interrupters in smart contracts will ensure the benefits for the smart contract owners. However, such a back-door mechanism may cause users to distrust the contracts. This worry makes sense, but we believe it can be fixed if the contract owners add some insurance mechanism to the contracts. For example, they can define rules to detect abnormal states, and the back-door mechanism can only be executed when the abnormal state is detected. For \textbf{``Missing Reminder”}, we did not receive comments from respondents who chose negative options. We sent emails to the developers who gave their email address and received three feedbacks. All mentioned that the smart contracts they developed are used inside their companies. They will write a detailed document of each function. If other developers in their companies have problems, then they fix the problems using face to face discussion. Therefore, this contract defect is not important for them. However, we believe that if the smart contracts are deployed on Ethereum and other developers can call the functions, removing this contract defect can reduce potential problems. For \textbf{``Unspecified Compiler Version"}, we found 4 developers who gave negative feedback mention that there are only very few differences between the versions under the same large version, e.g., between 0.4.21 and 0.4.22. However, we do not agree with this observation. As we have mentioned, even if two versions only have a small difference, but it is hard to foresee language constructions in the future version. Thus, it is possible that there might be two versions that contain a big difference in the future. Besides, refuting this feedback, version 0.4.0 (the first version of 0.4+) and 0.4.25 (the latest version of 0.4+) do indeed have big differences, as many APIs like \textit{throw} have been deprecated.

We also received 18 negative comments for the other 7 smart contract defects. The negative comments of \textbf{``Unmatched type assignment", ``Re-entrancy", ``Hard Code Address", ``Misleading Data location"}, and \textbf{``High Gas Consumption Function Type" } all mentioned that these contract defects have been removed in the latest version of Solidity. However, when developers deploy smart contracts to Ethereum, they need to choose a Solidity version by themselves. Most developers choose old versions of Solidity instead of the latest version~\cite{EtherScan_verify}. This means that these defects are still potentially harmful. \textbf{``Strict Balance Equality"} received 3 negative comments. Two developers said this is not a common case, and another developer said receiving Ether cannot be prevented. Thus, it might be hard to avoid exact balance checks in some situations.  We admit that defect is not common in Ethereum smart contracts. However, this defect is still harmful and can open up another attack vector to attackers. Developers can use other logic, such as ``$\geq$ \&\& \textless" to avoid ``=="  (see possible solution for this defect introduced in Section 3.3.1).  \textbf{``Unmatched ERC-20 standard"} received 2 negative comments. These comments mentioned that this contract defect could only be used for ICO smart contracts, which limits its usage scenario. However, ICO smart contracts are very popular in Ethereum, and they hold a large amount of Ethers. Thus, we I believe this defect category is still useful.

Certainly, We receive many positive comments. Some positive comments we received included:
\begin{itemize}
	
	\item  \emph{You provide \textbf{a very good summary} of some \textbf{very important security checkpoints}.}  
	\item	\emph{Those controls and warnings should be \textbf{integrated into the Solidity compiler}, and displayed in common development tools like Remix and Truffle.}
	\item  \emph{It is nice to have such a summary of these vulnerabilities among smart contracts, I think it would be \textbf{very helpful} for the blockchain practitioners as well as the researchers. }
	
	\item  \emph{These suggestions above are very useful to \textbf{avoid various kinds of flaws}. }
	
	\item  \emph{Generally speaking, all of these contract defects can lead to serious problems. \textbf{I learned a lot} from this survey.}
	
\end{itemize} 	


\section{RQ3: Distribution and Impact of Contract Defects}

\subsection{Motivation}
To help developers and researchers better understand the impacts of our defined smart contract defects, we summarized 5 impacts and manually label 587 smart contracts to show their distribution in the real-world smart contracts. Our labeling results provided ground truth for future studies on smart contract defects detection. As it is not easy to remove all contract defects due to tight project schedules or financial reasons, the impacts and distributions of different contract defects can help developers decide which defect should be fixed first.

\subsection{Approach}

\noindent  \textbf{Distribution:}
We obtained 587 smart contracts from real-world Ethereum accounts. The first and last authors of this paper independently read these smart contracts and determined whether the contracts contained our defined contract defects. They each have three-year experience on smart-contract-based development and have published three smart-contract-related papers together.  Their overall Kappa value was 0.71, which indicates substantial agreement between them. After completing the labeling process,  they discussed their disagreement and gave a final result. Finally, we generate a dataset which shows the distribution of the contract defects we defined. 

\noindent \textbf{Impact Level Definition:}
To summarize the impacts of each contract defect, we consider from three dimensions, i.e., contract dimension (unwanted behavior), attacker dimension (attack vector), and user dimension (usability), which can be found on Table~\ref{tab:impactFeature}. 

The contract dimension focuses on the severity level of the contract defect. From our survey, 27 developers claimed that defects, e.g., \textit{Reentrancy, Dos Under External Influence}, might enable attackers to attack the contracts, and 9 of them mentioned that attackers can utilize defects like \textit{Reentrancy} to stole all the Ethers on the contract. Also, 16 developers agree that defects, e.g., \textit{High Gas Consumption Function Type, Deprecated APIs}, will not affect the normal running of the contract, but have bad effects for the users or callers. 
From the \textit{StackExchange} posts, we can also find the comments of the posts mentioned that the defects could lead to the crashing, losing all Ethers, and losing a part of the Ethers.  Finally, we totally find the defects can lead to 5 common consequences to the contracts. They are crashing, being controlled by attackers, losing all Ethers, losing a part of the Ethers, normal running but have bad effects for the users or caller. We have split the 5 common consequences into three severity levels, i.e., \textit{critical}, \textit{major}, and \textit{trivial}. \textit{Critical} represents contract defects, which can lead to the crashing, being controlled by attackers, or can lose all Ethers. \textit{Major} represents the contract defects that can lead to the loss of a part of the Ethers. Contracts with \textit{trivial} severity level will not affect the normal running of the contract.

The attacker dimension focuses on attackers’ behaviors. Since financial services are the most attractive targets for attackers, we believe that if attackers can use the defects to steal Ethers, the impact level should be higher. Whether the defect can be triggered by attackers is also an important aspect.

The users dimension focuses on the external influence of the defects. This dimension contains three aspects, i.e., potential errors for caller, gas waste, and mistakes on code reuse. Some defects do not affect the normal running of the contracts. However, they can lead to the errors of the caller programs. Some defects can also increase the gas costs of the callers and users. As code reuse is important in software engineering, some defects can make the contracts hard to be understand and reuse.

We only consider the worst-case scenario outcome for each contract defect, even though some defects will have different impact levels under different application scenario. We use \textit{Hard Code Address} as an example. In most situations, Hard Code Address will not lead to the loss of Ethers. However, if the hard-coded address is a self-destructed contract, a contract with this defect can lose a part of its Ethers. Thus we consider \textit{Hard Code Address} can lead to major unwanted behavior.

After defining the three dimensions, we map each contract defect onto one or more. The detailed results are shown in Table. We found there are 5 common types of distribution. According to the distribution, we summarized 5 impact levels and assigned each contract defect to have one impact level.

\begin{table*}[htbp]
	\scriptsize	
	\centering
	\caption{Features of Each Contract Defects}
	\begin{tabular}{c|c | c| c |p{40pt} <{\centering} | p{20pt} <{\centering} |p{45pt} <{\centering} | c | p{35pt} <{\centering}}
		\hline
		\multirow{2}[4]{*}{Contract Defects} & \multicolumn{3}{c|}{Unwanted Behavior} & \multicolumn{2}{c|}{Attack Vector} & \multicolumn{3}{c}{Usability} \\
		\cline{2-9}          & Critical & Major & Trivial & Triggered by External & Stolen Ethers & Potential Errors for Callers & Gas Waste & Mistakes on Code Reuse  \\
		\hline
		Unchecked External Calls   &      &  \checkmark     &       &      &      &       &       &  \\
		\hline
		Dos Under External Influence   & \checkmark     &       &       & \checkmark     &       &       &       &  \\
		\hline
		Strict Balance Equality  &  \checkmark     &      &       &   \checkmark    &       &       &       &  \\
		\hline
		Unmatched Type Assignment   &    \checkmark   &       &      &    \checkmark   &       &     &       &  \\
		\hline
		Transaction State Dependency   &    \checkmark   &       &     &    \checkmark   &   \checkmark    &       &      &  \\
		\hline
		Reentrancy   &   \checkmark    &       &     &   \checkmark    &   \checkmark    &       &      &  \\
		\hline
		Hard Code Address   &       &   \checkmark    &     &       &       &       &      &  \\
		\hline
		Block Info Dependency   &       &   \checkmark    &     &   \checkmark    &       &       &      &  \\
		\hline
		Nested Call   &   \checkmark    &       &     &   \checkmark    &       &       &      &  \\
		\hline
		Deprecated APIs   &       &       &  \checkmark    &       &       &       &  \checkmark    & \checkmark  \\
		\hline
		Unspecified Compiler Version   &       &       &  \checkmark    &       &       &       &    \checkmark  & \checkmark \\
		\hline
		Misleading Data Location   &   \checkmark    &       &     &    \checkmark   &       &       &      &  \\
		\hline
		Unused Statement   &       &       &   \checkmark   &       &       &       &   \checkmark   & \checkmark \\
		\hline
		Unmatched ERC-20 standard   &       &       &  \checkmark    &       &       &   \checkmark    &      &  \\
		\hline
		Missing Return Statement   &       &       &  \checkmark    &       &       &       &   \checkmark   & \checkmark \\
		\hline
		Missing Interrupter   &       &       &  \checkmark    &       &       &    \checkmark   &      &  \\
		\hline
		Missing Reminder   &       &       &  \checkmark    &       &       &   \checkmark    &      &  \\
		\hline
		Greedy Contract  &    \checkmark   &       &     &       &       &       &      &  \\
		\hline
		High Gas Consumption Function Type   &       &       &   \checkmark   &       &       &       &    \checkmark  &  \checkmark\\
		\hline
		High Gas Consumption Data Type  &       &       &  \checkmark    &       &       &       &   \checkmark   & \checkmark \\
	\end{tabular}%
	\label{tab:impactDefect}%
\end{table*}%


\subsection{Results}

\begin{table*}
  \centering
  \scriptsize	
  \caption{Features of Each Impact Level}
    \begin{tabular}{c|c c c|c c|c c c}
    \hline
    \multirow{2}[4]{*}{Impact Level} & \multicolumn{3}{c|}{Unwanted Behavior} & \multicolumn{2}{c|}{Attack Vector} & \multicolumn{3}{c}{Usability} \\
\cline{2-9}          & Critical & Major & Trivial & Triggered by External & Stolen Ethers & Potential Errors for Callers & Gas Waste & Mistakes on Code Reuse  \\
    \hline
    IP1   & \checkmark     &       &       & \checkmark     & \checkmark     &       &        &   \\
    \hline
    IP2   & \checkmark     &       &       & \checkmark     &       &       &       &  \\
    \hline
    IP3   &   T1    &  T2     &       &    T2   &       &       &       &  \\
    \hline
    IP4   &       &       & \checkmark     &       &       & \checkmark    &       &  \\
    \hline
    IP5   &       &       & \checkmark    &       &       &       & \checkmark     & \checkmark \\
    \end{tabular}%
  \label{tab:impactFeature}%
\end{table*}%

We use Table~\ref{tab:impactFeature} to clarify the difference between each impact level. IP1 is the highest, and IP5 is the lowest.  Contract defects with impact level 1-2 can lead to critical unwanted behaviors, like crashing or a contract being controlled by attackers. Contract defects with impact level 3 can lead to major unwanted behaviors, like lost ethers. Impact level 4-5 can lead to trivial problems, e.g., low readability, which would not affect the normal running of the contract. 

The detailed definition of the five impact levels are as follows:

\noindent\textbf{Impact 1 (IP1)}: The smart contracts containing the related contract defects can lead to critical unwanted behaviors. Unwanted behaviors can be triggered by attackers, and they can make profits by utilizing the defects.

\noindent\textbf{Impact 2 (IP2)}: The smart contracts containing the related contract defects can lead to critical unwanted behaviors. Unwanted behaviors can be triggered by attackers, but they cannot make profits by utilizing the defects.

\noindent\textbf{Impact 3 (IP3)}: There are two types of IP3. \textit{Type 1:} The smart contracts containing the related contract defects can lead to critical unwanted behaviors, but unwanted behaviors cannot be triggered externally. \textit{Type 2: }The smart contracts containing the related contract defects can lead to major unwanted behaviors. The unwanted behaviors can be triggered by attackers, but they cannot make profits by utilizing the defects.

\noindent\textbf{Impact 4 (IP4)}:  The smart contracts containing the related contract defects can work normally. However,  the contract defects can lead to potential risks of errors when outside programs call the contracts.

\noindent\textbf{Impact 5 (IP5)}: The smart contracts containing the related contract defects can work normally and will not lead to the errors for the callers. However, the contract defects can lead to gas waste, and make the contracts hard to understand and reuse. 

Table~\ref{tab:result} lists the detailed distribution of each contract defect (the fourth column) in our dataset and its related impact (the last column). We find the distribution for Impacts 1 -- 5 to be 2.90\%, 7.16\%, 27.09\%, 93.86\%, 99.14\%, respectively. Note that one smart contract can have multiple defects of different impacts simultaneously. 

\textit{``Unspecified Compiler Version”} is the most common contract defect in our dataset (90.63\% contracts contain this defect). We also found that this contract defect is the most popular one among the 20 defects when we analyze StackExchange posts. Many developers want to reuse the contracts but encounter compiler errors. These contracts usually do not specify a compiler version. In this case, developers have to try different compiler versions or refactor the code, which increases the workload for code reuse.

\textit{``Missing Interrupter”} is also very popular in our dataset (89.1\% contracts contain this defect). This defect receives the greatest number of comments in our the survey. On the one hand, developers admit that adding interrupter is important for contracts when emergencies happen. On the other hand, some developers also worried that the interrupter could lead to distrust by the contract users. Better understanding attitudes to this defect may need further research effort. For example, researchers can design a survey for developers to investigate the reasons why they add or do not add interrupters. By knowing the reason why developers do not add it, researchers might design a better method to implement interrupter. By knowing the reason why developers add interrupters, researchers can investigate whether contracts with interrupters in our dataset are consistent with these reasons, and what are the most popular reasons.

99.82\% of smart contracts in our dataset contain at least one contract defect of the impact 4 or impact 5. These contract defects will not affect the normal running of the contracts, but it may have unpredictable impacts to the caller or code reuse. The distribution may illustrate that the developers focus more on the functionality but do not consider the code reuse or handle unpredictable behaviors caused by attackers. This finding is similar to Chen et. al~\cite{chen2018understanding}. They found that 96\% of smart contracts are involved in no more than 5 transactions, and they are not be used anymore, indicating that many developers do not consider future reuse of these contracts.

About 32.03\% of smart contracts contain contract defects at levels 1-3, which can lead to unwanted behaviors. However, we found that only 7.33\% of smart contract contains defects that can lead to critical unwanted behaviors, e.g., crashing or being controlled by attackers.

We also found that ERC-20 related smart contracts are the most popular (36.11\%) in Ethereum. However, 21.22\% of them do not strictly follow the ERC-20 standards. We did not find any smart contracts which contain \textit{High Gas Consumption Data Type}. Since the size of our dataset is limited, and this contract defect has related posts on \textit{StackExchange}, this contract defect might exist if we investigate more contracts. In summary, our findings showed that defined contract defects are very common in real-world smart contracts.


\section{Discussion}
In this section, we first give the implications of our work for researchers, practitioners and educators. Then, we list three challenges for future research on automatic contract defect detection. Finally, we give the threats of validity. 

\subsection{Implications}
\noindent \textbf{For Researchers:} \textit{Research Guidance.} In this paper, we defined 20 contract defects. Several previous studies analyzed some of them. We have investigated whether there are existing tools that can detect some of the contract defects identified by our work. We show the results in Table~\ref{tab:Tools}. We first collected the titles of papers which were published at CCS, S\&P, USENIX Security, NDSS, ACSAC, ASE, FSE, ICSE, TSE, TIFS, and TOSEM from 2016 to 2019, since Ethereum went live on July 30, 2015~\cite{eth_Introduction}. Then, we used the keywords ``smart contract", ``Ethereum", ``blockchain", ``Contracts" to search for papers which are related to the smart contract technology. After that, we read the abstract of each paper to verify its relevance. Finally, we found a total of 4 related papers (i.e., Oyente~\cite{oyente,oyente-tools} , Zeus~\cite{Zeus}, Maian~\cite{Maian} and Contractfuzzer ~\cite{Contractfuzzer}). We provide a description of these four tools in Section~\ref{related}. We find that 7 contract defects can be detected by these existing tools and most of them are security related defects.  These tools focus more on the security aspects but do not consider the other two aspects considered as equally important by practitioners. Therefore, researchers can pay more attention to developing tools that can detect the other 13 contract defects.

\noindent\textit{Behavior vs. Perception~\cite{devanbu2016belief}.}
The belief of whether a contract defect is important or not may result in prioritizing testing effort. The survey results and contract defect distribution shown in Table~\ref{tab:result} can help us investigate whether the practitioners' perception is consistent with their behavior. We  find that the top two most frequent contract defects are `Unspecified Compiler Version' and `Missing Interrupter' (according to the column \emph{No. Defects} in Table~\ref{tab:result}). Their survey scores are also the lowest (3.92 and 4.0 according to the column \emph{Score} in Table~\ref{tab:result}), indicating that practitioners do not perceive them as important as other defects, and thus they pay less attention to them in practice which causes them to appear more than other contract defects. The appearance of these two contract defects is consistent with practitioners' perception. However, there are many inconsistent examples. According to the definition of 5 impacts introduced in Section 5.3, it is clear that IP1 can cause the most serious problems compared to other impacts. We find the `Unchecked External Calls' has the second highest survey score (4.64), which shows that developers think this defect is very important.  However, its impact is IP3, which shows that there is an inconsistency between the practitioners' perception (high survey score) and their behavior (medium impact to the project). Future contract defect detection tools should provide rationales that explicitly describe the connection between contract defects and its impact. This could assist developers better prioritize testing efforts, and understand the detection results well.

\noindent\textit{Contract Defects in Other Smart Contract Platforms. }We propose a method which summarizes contract defects from online posts. Our study focused on defining contract defects for Ethereum smart contracts, but the same method can be applied to other popular blockchain platforms, e.g., EOS~\cite{EOS}, Hyperledger~\cite{Hyperledger}. These blockchain platforms also support the running of smart contracts and have their unique features. There are thousands of posts on \textit{StackExchange} related to these platforms. Researchers can analyze the related posts and find specific features and contract defects of these smart contract platforms. Our work defined 20 contract defects and provide a dataset which identifies these contract defects on 587 contract accounts, which point out a new direction for future research. For example, researchers can develop automatic contract defect detection tools, and our dataset can be used as ground truth to validate the performance of these tools.

\begin{table} 
	\scriptsize
	\caption{Tools that  detect some contract defects identified by our study.}
	\label{tab:Tools}
	\centering
	\begin{tabular}{c | c  }
		\hline
		Contract Defects & Tools\\
		\hline
		Unchecked External Calls & Oyente, Zeus, Contractfuzzer\\
		Reentrancy & Oyente, Zeus, Contractfuzzer\\
		Block Info Dependency& Oyente, Zeus, Contractfuzzer\\
		Transaction State Dependency& Oyente, Zeus\\
		DoS Under External Influence& Zeus \\
		Unmatched Type Assignment& Zeus\\
		Greedy Contract& Maian \\
		\hline
	\end{tabular}	
\end{table}

\noindent \textbf{For Practitioners:}
We are the first to conduct an empirical study by analyzing many online StackExchange posts to understand and define contract defects for smart contracts, and utilize an online survey to validate the acceptance of the defined contract defects among real-world developers. Our results showed that most of the smart contracts in our dataset contained at least one of the defined contract defects. The results may indicate that developers do not consider future use and handle unpredictable attacks. However, since the smart contracts are immutable to patch, the consideration of future use and unpredictable attacks is very important. We also concluded 5 impacts of the defined contract defects to help practitioners better understand the consequences. The defined contract defects can be regarded as a coding guidance for practitioners when they develop smart contracts.  By removing the defined contract defects, they can develop robust and well-designed smart contracts.

Developing contract defect detection tools is also a good direction. Our online survey received many comments from managers of smart-contract-related companies, some listed in Section 4.3. They showed much interest in developing and using related tools and highlighted that such detection tools should be integrated into Solidity compiler and development tools.

\noindent \textbf{For Educators:}
Educators should emphasize the importance of removing contract defects before deploying smart contracts to blockchain. A survey~\cite{blockchainnews} shows that more than 20\% of top 50 universities are offering blockchain courses until Oct. 2018. However, most courses focus on teaching basic grammar rule of Solidity programming or blockchain related knowledge but ignore other concerns (security, architecture, usability). The distribution of the defined contract defects also indicates that many developers do not realize the importance for the reuse of smart contracts and handling unpredictable attacks. Educators can improve such conditions by helping students to better understand the impacts of the contract defects. Thus, it is highly recommended that educators pay more attention to teaching contract defect related problems for smart contract development.

\subsection{Challenge in Detection Contract Defects}
We point out three challenges to give a guideline for future research on automatic contract defect detection.

\noindent \textbf{(1) Program Understanding. }
Some contract defects do not have a specific pattern, which increase the difficulty of automatic defection. For example, there are multiple methods to implement interrupter for the contracts. Developers can use \textit{selfdestruct} function to kill the contract. They can also write a method to stop the contract when attack happens. To detect these kinds of contract defects, we need to understand the smart contracts. However, automatically understanding code is not easy.

\noindent \textbf{(2) Bytecode Level Detection. }
When deploying a smart contract to Ethereum, EVM will compile the source code to the bytecode and the bytecode will be stored on the blockchain. Everyone can check the bytecode of the smart contracts, but  source code may not visible to the public. Smart contracts usually call other contracts, but the callee contracts may not open their source code to inspection. In other words, they do not know whether the smart contract they called is safe or not. Therefore, detecting contract defects through bytecode is very important because each smart contract’s bytecode can be found on Ethereum but only around 0.45\% of smart contracts have opened up their source code by Jan. 2019~\cite{EtherScan}. However, it is not easy to detect contract defects from bytecode level as it loses the most semantic information. 

\noindent \textbf{(3) EVM Operation. }
When compiling a smart contract to bytecode, EVM will optimize the source code, which means some information will be removed or optimized, so it is hard to know the original information on the source code. For example, detecting whether a function has return value on source code level is straightforward. However, it is not easy to detect it at bytecode level as even we do not add a return value for a function, the EVM will add a default value for it. Therefore, we cannot know whether the return value is added by EVM or developers.

\subsection{Possible Detection Methods}
In this section, we discuss possible detection methods for each of the contract defects that we have defined. Since 7 defects shown in Table~\ref{tab:Tools} have already been detected by previous tools, we only discuss the remaining 13 defects.

\subsubsection{Bytecode Level Detection}
Detecting contract defects by bytecode is important for smart contracts in Ethereum, as all the bytecode of the contracts can be found on the Ethereum, but only less than 1\% contracts have open source code. To detect contract defects by bytecode, the defects should have regular patterns. For example, \textit{Nested Call} can be found in a loop which does not limit its loop times and contains the \textit{CALL} instruction. \textit{Missing Interrupter }does not have a regular pattern, as there are multiple ways to realize interrupter. To the best of our knowledge, we have found 6 contract defects that can be detected by bytecode among our 13 smart contract defects. A common method to detect defects by bytecode is using symbolic execution as it can statically reason about a program path-by-path~\cite{oyente}.. The method usually converts bytecode to the opcode and splits them into several blocks.\footnote{A basic block is a straight-line code sequence with no branches in except to the entry and no branches out except at the exit.} A basic block is a straight-line code sequence with no branches in except to the entry and no branches out except at the exit. Then, we can symbolically execute the instruction and construct a control flow graph (CFG) for each contract, which can be used to detect the contract defects.

\textbf{(1) Nested Call:}  After obtaining the CFG, we can identify which blocks belong to loops. If the loop body contains CALL instructions and does not limit its loop iterations, the loop contains a \textit{Nested CALL }defect.

\textbf{(2) Strict Balance Equality:} To get the balance of the contract, the contract will generate a \textit{BALANCE} instruction. We can start from this instruction; If a \textit{BALANCE} instruction is read by EQ (the EQ instruction is used to compare whether two values are equal), it means there is a strict balance equality check. If this check happens at a conditional jump expression, it means this contract contains a Strict Balance Equality defect.

\textbf{(3) Hard Code Address:} Addresses of Ethereum strictly follow the EIP55~\cite{EIP55} standard. We need to identify whether the opcode contains a 20-byte-value and follow the EIP55 standard. The default bytecode stored on Ethereum is called runtime bytecode, which does not contain the constructor function. However, many hard code addresses are stored in the constructor function. To obtain the constructor function, we can check the value of the first transaction of the contract.

\textbf{(4) Unmatched ERC-20 standard:} The ERC-20 standard contains 9 functions (3 are optional). From bytecode, we can get the hash value of each function. The hash value is obtained from its function name and parameter types. For example, the hash value of ``transfer (address, uint256)” is ``A9059CBB". Therefore, we can identify whether a contract is an ERC20 token contract by comparing the hash value of each function. Then, we need to check whether each function strictly follows the ERC-20 standard.

\textbf{(5) High Gas Consumption Function Type:} We can identify the public functions through CFG. If a public function is not be called by any other function this means the function can be changed to an ``external” function.

\textbf{(6) High Gas Consumption Data Type:} To detect this defect, we need to identify the pattern of byte[] from opcode. byte[] is easy to identify as it always occupies multiples of 32 bytes.

\subsubsection{Source Code Level Detection}
As we introduced in Section 6.2 (3), a part of the information will be removed or optimized when compiling the source code to the bytecode. Therefore, the remaining 7 contract defects need to be detected from smart contract source code.

\textbf{(1) Deprecated APIs:} Solidity document does not suggest using some APIs in the latest version, as they will be deprecated in the future. However, these APIs can still be compiled.  When compiling to the bytecode, their instructions might be the same as the recommended APIs. To detect deprecated APIs, we need to use the latest version of Solidity and detect which APIs are deprecated. 

\textbf{(2) Unused Statement:} Since some unused statements will be optimized by the EVM, this defect should be detected from source code. To detect this defect, we can compile the contract by using the Solidity compiler~\cite{solc}  and compare it to the original contract. There might be some unused statements that cannot be optimized by EVM. To detect these unused statements, we can utilize the CFG and detect whether all the paths can be executed.

\textbf{(3) Unspecified Compiler Version:} When compiling source code to bytecode, developers need to choose a specific version of the Solidity compiler. In this case, we cannot detect the defect from its bytecode. To detect this defect, we need to check its pragma solidity.~\cite{Solidity}

\textbf{(4) Misleading Data Location:} If a smart contract has a \textit{Misleading Data Location}, we will find that the contract modified the value on a specific storage position. However, we cannot know whether this operation is due to the contract defect. In this case, we need to detect the defect from source code. To detect it, we first need to check whether there is an array, struct, or mapping created in a function. Then, if the contract pushes a value before assigning to a storage value, this defect is detected.

\textbf{(5) Missing Return Statement:}  The reason for this has been introduced in Section 6.2.3. To detect this defect, we can split source code into functions by using AST (abstract syntax tree), and check whether a function is missing a return statement.

\textbf{(6) Missing Interrupter:} There are multiple ways to realize interrupters, so we cannot find a method to detect this defect from bytecode. To detect the defect, we first need to summarize common methods of realizing interrupters. Then, we detect each kind of interrupter. For example, adding a \textit{selfdestructor} function is one of the interrupters. In this case, we just need to detect whether a contract contains a \textit{selfdestructor} function.

\textbf{(7) Missing Reminder:} There are also many kinds of functions that need to add reminders. To detect this defect, we all need to summarize what kind of functions need to add a reminder, then detect the defect one by one. For example, when receiving Ethers, we might use a reminder to throw an event to inform the user. In this case, we first need to locate function that can receiving Ethers. Then, verifying whether the function throws an event to inform users.

\subsection{Code Smells in Ethereum}

In software engineering, code smells are the symptoms in the source code that possibly indicate deeper problems~\cite{smellDefinition}. Code smells are related to not only security issues but also design flaws, which might slow down development or increase the risk of bugs or failures in the future. Detecting and refactoring out code smells helps increase software robustness and enhance development efficiency~\cite{van2002java}. In this paper, we defined 20 contract defects. There are many similarities between code smells and contract defects.  According to Martin Fowler’s book~\cite{smellDefinition}, code smells do not directly trigger bugs but can lead to ``potential'' program faults. This definition is similar to the definition of Impact level 4 and 5. According to our definition, the contracts containing contract defects with impact level 4 and 5 can work normally, but they can lead to potential risks of errors when outside programs call the contracts, or increase the difficulty of code reuse. In this case, the contract defects with IP4 and IP5, e.g., ``\textit{Unused Statements}'', ``\textit{Unspecified Compiler Version}'', can also be considered as smart contract code smells.

\section{Threats to Validity}
\subsection{Internal Validity}
We used keywords to filter StackExchange posts. The scale of our keywords dataset determines how much manual effort we need to pay. It is not easy to cover all keywords, which means we may not cover all contract defects. Due to the time and human resource limitation, we defined 20 contract defects in this study, but researchers can define more contract defects by using our methods. To reduce this threat, we manually labeled 587 smart contracts to validate the existing of these contract defects. To provide a more stable labeling process, we followed the card sorting process, and two authors labeled the smart contracts independently. However, it is still possible that some errors exist in our dataset because of misunderstanding of smart contracts. To reduce the errors, we choose the most experienced authors to label the contracts. They each have three-year experience on smart contract based development and have published several smart contract related papers. 

The impact of smart contract defects depend on our understanding of each contract defect. However, different researchers and developers may have different understandings. To minimize this threat, we read the related posts and real-world examples and discussed with several smart contracts developers to help improve the correctness. We also considered feedback and comments from our survey.


It is difficult to ensure that all developers have a good understanding about all of the contract defects and are indeed paying attention when doing our survey. It is possible that some feedback might contain incorrect information. For example, some survey respondents give ``very important" or ``very unimportant" feedback to all defects. To reduce the influence of this situation, we first added an option ``I don't understand" to each question and removed these responses when analyzing our survey data.  We also made each question optional. Therefore, if developers find that a question is hard to understand or they lose their patience, they can skip the question instead of giving incorrect answers. Finally, we remove feedbacks given by developers whose answers are all the same when analyzing the survey data, e.g., all ``very important", all "very unimportant".  In addition, to help Chinese developers better understand our contract defects, three Chinese authors of this paper translated the survey into Chinese and reviewed the translated version to make sure the translation is correct. 

\subsection{External Validity}  
Solidity is a fast-growing programming language. In 2018, 9 versions were updated and released~\cite{solidityVersion}, which means many features may be added or removed in the future. Ethereum can also be updated through hard fork~\cite{hardfork}. The latest hard fork named Constantinople will happen on the first half of 2019~\cite{ethereum}. Constantinople will add five new Ethereum Improvement Proposals (EIPs) to ensure proof-of-work more energy efficient. Some new opcodes will be added (e.g., CREATE2) and some opcodes will be modified (e.g., SSTORE). This means some new contract defects may be created, or existing contract defects will be modified. Thousands of new smart contracts may quickly be deployed to the blockchain. The distribution of the contract defects on real-world smart contracts may change with new developments of smart contract technology. Many new posts are uploaded to the \textit{StackExchange}, and these posts can expose new contract defects. Our method can also be applied to this situation, but it needs further effort. 
\section{Related Work}
\label{related}

Atzei et al.~\cite{atzei2016survey} proposed the first systematic exposition survey on attacks on Ethereum smart contracts. They introduce 12 kinds of security vulnerabilities from Solidity, EVM, and Blockchain level. Besides, they also introduce some attacks, which can be used by the attackers to make profits. The work claims that security vulnerabilities introduced in the paper are obtained from academic literature, Internet blogs, discussion forums, and based on authors' practical experience on programming smart contracts. However, the paper does not introduce the detailed steps of finding the vulnerability and does not validate whether developers consider these vulnerabilities as harmful. Another difference with our work is that our work does not only focus on  the security aspect. Instead, we consider from security, availability, performance, maintainability and reusability aspects.

Oyente~\cite{oyente,oyente-tools} is the first bug detection tool of smart contracts, which utilizes symbolic execution to detect four security issues, i.e., mishandled exception, transaction-ordering dependence, timestamp dependence and reentrancy attack. First, Oyente builds a skeletal control flow graph for the input contracts. Then, they faithfully simulate EVM code and execute the instructions to produce a set of symbolic traces. After that, Oyente defines different patterns to check whether the tested contracts contain the security problems or not. Oyente measured 19,366 existing Ethereum contracts and found 8,519 of them contain the defined security problems.

Kalra et al.~\cite{Zeus} found many false positives and false negatives in Oyente's results. They developed a tool called Zeus, an upgraded version of Oyente. Their tool feeds Solidity source code as input and translates them to LLVM bitcode. Zeus detects 7 security issues, 4 of them are the same as Oyente and other 3 problems are \emph{unchecked send, Failed send, Integer overflow/underflow}. To evaluate their tool, Kalra crawled 1524 distinct smart contracts from Etherscan~\cite{EtherScan}, Etherchain~\cite{EtherChain} and EtherCamp~\cite{EtherCamp} explorers. The result indicates about 94.6\% of contracts contain at least one security problem.

Jiang et al.~\cite{Contractfuzzer} focus on 7 security vulnerabilities, i.e., Gasless Send, Exception Disorder, Reentrancy, Timestamp Dependency, Block Number Dependency, Dangerous DelegateCall and Freezing Ether.  They also developed a tool named ContractFuzzer to detect these issues. Their tool consists of an offline EVM instrumentation tool and an online fuzzing tool. Based on smart contract ABI, ContractFuzzer can automatically generate fuzzing inputs to test the defined security issues. They tested 6,991 smart contracts and found that 459 of them have vulnerabilities.

Nikolic~\cite{Maian} et al. focus on security issues that can lead to a contract not able to release Ethers, can transfer Ethers to arbitrary addresses, or can be killed by anybody. Their tool, MAIAN, takes as input data either Bytecode or source code. MAIAN contains two major parts: symbolic analysis and concrete validation. Like Oyente, simulates an Ethereum Virtual Machine, utilizes symbolic execution, and defines several execution rules to detect these security issues. Their results were deduced from 970,898 smart contracts and found that a total of 34,200 (2,365 distinct) contracts contain at least one of these three security issues.

Gao~\cite{smartembed}  et al. designed a tool named \textsc{SmartEmbed}, which detect bugs in smart contracts by using a clone detection method.  \textsc{SmartEmbed} contains a training phase and a prediction phase. In the training phase, there are two kinds of  dataset, i.e.,  source code database and bug database. Source code database contains all the verified (open sourced) smart contracts in the Etherscan. The bug database records the bugs of each smart contract in their source code database. To build the prediction modle,  \textsc{SmartEmbed} first converts each smart contract to an AST(abstract syntax tree). After normalizing the parameters and irrelevant information on the AST, \textsc{SmartEmbed} transfers the tree structure to a sequence representation. Then, they use \textit{Fasttext}~\cite{bojanowski2017enriching} to transfer code to embedding matrices. Finally, they compute the similarity between the given smart contracts with contracts in their database to find the clone contracts and clone related bugs.

We defined 20 contract defects from three different aspects. The above four papers introduce some security problems while we focus on a broader problem coverage. We do not just focus on security problems but help developers build better smart contracts. We also define patterns to help developers increase software usability and architecture. While these works show several security problems, but did not validate whether practitioners consider these problems as harmful. Our work not only validated our defined defects by an online survey, but also analysis their impacts and distribution, which can give a clear guidances for developers.

\section{Conclusion and Future work}
We conducted the first empirical study to understand and characterize smart contract contract defects. We first selected 4,141 warning related StackExchange posts from 17,128 posts. Then we manually analyzed these posts and defined 20 smart contract defects from five aspects -- security, availability, performance, maintainability and reusability problems. To validate our defined contract defects, we created an online survey. The feedback from our survey indicates our contract defects are important and addressing them can help developers improve the quality of their smart contracts. We analyzed the impacts for each contract defect and labeled 587 real-world smart contracts from Ethereum platform. 

Two groups can benefit from this study. For smart contract developers, they can develop more robust and better-designed smart contracts. The 5 impacts could help developers decide the priority of removal. For software engineering researchers,  our dataset can provide ground truth for them to develop smart contract defect detection tools. We plan to develop automated contract defect detection tools to detect these defined contract defects. We also plan to extend our contract defect list and dataset, when more posts will be published in StackExchange, and more features will be added into \textit{Solidity} in the future.

\noindent \textbf{Acknowledgment.} This research was partially supported by the Australian Research Council’s Discovery Early Career Researcher Award (DECRA) funding scheme (DE200100021), ARC Discovery Project scheme (DP170101932), ARC Laureate Fellowship (FL190100035), Hong Kong RGC Project (No. 152193/19E), National Natural Science Foundation of China (61872057), and National Key R\&D Program of China (2018YFB0804100)

\end{document}